\documentclass[journal]{IEEEtran}
\usepackage{graphicx}
\usepackage[normalem]{ulem}

\usepackage[swedish,english]{babel}

\usepackage{tikz,amsmath,amssymb,amsfonts,bm}
\usepackage{pgfplots}
\pgfplotsset{compat=newest}

\newcommand{\G}{{G}}

\newcommand{\vr}{\vec{r}}
\newcommand{\vJ}{\vec{J}}

\renewcommand{\vec}[1]{{\boldsymbol#1}}
\newcommand{\diff}{\mathrm{d}}
\providecommand*{\mrm}[1]{\mathrm{#1}}
\providecommand*{\vhat}[1]{\hat{\vec{#1}}}
\providecommand*{\iu}{\ensuremath{\mrm{j}}}
\providecommand*{\eu}{\ensuremath{\mrm{e}}}
\renewcommand{\Im}{\mathop{\mathrm{Im}}}
\renewcommand{\Re}{\mathop{\mathrm{Re}}}
\newcommand{\Om}{\Omega}
\newcommand{\ie}{\textit{i.e.}\/ }

\newcommand{\R}{\mathbb R}

\newcommand{\Z}{\mathbb Z}
\newcommand{\C}{\mathbb C}

\newcommand{\We}{W_{\rm e}}
\newcommand{\Wm}{W_{\rm m}}
\newcommand{\Webf}{{\rm \bf W}_{\rm e}}
\newcommand{\Wmbf}{{\rm \bf W}_{\rm m}}
\newcommand{\Wbf}{{\rm \bf W}}
\newcommand{\Walpha}{{\rm \bf W}_{\alpha}}
\newcommand{\Xbf}{{\rm \bf X}}
\newcommand{\Rbf}{{\rm \bf R}}
\newcommand{\Ibf}{{\rm \bf I}}
\newcommand{\Abf}{{\rm \bf A}}
\newcommand{\herm}{\mrm{H}}
\newcommand{\minimize}{\mathop{\mrm{minimize}}}
\newcommand{\subto}{\mrm{subject\ to}}
\newcommand{\trace}{\mathop{\mrm{tr}}}
\newcommand{\Kzmn}{k_{{\rm z}mn}}
\newcommand{\Kxm}{k_{{\rm x}m}}
\newcommand{\Kyn}{k_{{\rm y}n}}

\newcommand{\Ktmn}{\vec{k}_{{\rm t}mn}}

\newcommand{\Ktnull}{\vec{k}_{{\rm t}00}}

\renewcommand{\Pr}{P_{\rm r}}
\newcommand{\Pohm}{P_{\rm Ohm}}

\newcommand{\diag}{{\rm diag}}

\DeclareMathAlphabet\mathbfcal{OMS}{cmsy}{b}{n}
\DeclareMathOperator{\rank}{rank}
\newcommand{\unitz}{\overline{\overline{\mathbf{1}}}_{\rm z}}
\newcommand{\unitdyadic}{\overline{\overline{\mathbf{1}}}}
\newcommand{\rade}{e_{\rm rad}}
\newcommand{\Hd}{\overline{\overline{\mathbf{H}}}}

\newcommand{\lx}{\ell_{\rm x}}
\newcommand{\ly}{\ell_{\rm y}}

\usepackage{systeme}
\usetikzlibrary{decorations}
\usetikzlibrary{decorations.markings}

\usepackage{cite}

\begin{document}

\title{Physical Limitations of Phased Array Antennas}

\author{Andrei Ludvig-Osipov, Jari-Matti Hannula, Patricia Naccachian, and B. L. G. Jonsson 
\thanks{This work was supported by the Vinnova Center under the ChaseOn/iAA project, the Swedish Foundation for Strategic Research (SSF) under the grant AM130011, and the Walter Ahlstr{\"o}m Foundation.}%
\thanks{ A.~Ludvig-Osipov was with the School of Electrical Engineering and Computer Science, KTH Royal Institute of Technology, Stockholm SE-10044, Sweden. He is now with the Department of Electrical Engineering, Chalmers University of Technology, SE-41296 G\"oteborg, Sweden (e-mail: osipov@chalmers.se).}
\thanks{J.-M.~Hannula, and B.~L.~G.~Jonsson are with the School of Electrical Engineering and Computer Science, KTH Royal Institute of Technology, Stockholm SE-10044, Sweden (e-mail: jari-matti.hannula@ieee.org; ljonsson@kth.se.).}
\thanks{P.~Naccachian is with the Department of Electrical and Computer Engineering, American University of Beirut, Beirut, Lebanon (e-mail: ppn02@mail.aub.edu)}
}

\maketitle

\begin{abstract}
In this paper, the bounds on the Q-factor, a quantity inversely proportional to bandwidth, are derived and investigated for narrow-band phased array antennas.
Arrays in free space and above a ground plane are considered. The Q-factor bound is determined by solving a minimization problem over the electric current density. The support of these current densities is on an element-enclosing region, and the bound holds for lossless antenna elements enclosed in this region.
The Q-factor minimization problem is formulated as a quadratically constrained quadratic 
optimization problem that is solved either by a semi-definite relaxation or an eigenvalue-based method.
We illustrate numerically how these bounds can be used to determine trade-off relations between the Q-factor and other design specifications: element form-factor, size, efficiency, scanning capabilities, and polarization purity.
\end{abstract}


\section{Introduction}
\label{sec:intro}

One of the most important antenna design parameters is the impedance bandwidth, for which an antenna satisfies its design criteria. Directly optimizing the bandwidth performance of an antenna or an array is difficult as it is closely related to the shape-optimization. In particular, it may be hard to determine if a global optimum has been reached. A class of tools addressing this problem are the fundamental bounds (limitations imposed by physical principles), and they have been instrumental for determining optimal performance for small antennas~\cite{Chu1948,Gustafsson+etal2007a,Gustafsson+tutorial2016}. 

In this paper, we use the  Q-factor~\cite{LudvigOsipov+Jonsson2019,LudvigOsipov+Jonsson2019gp} to determine bounds for narrow-band phased array antennas in free space and above a ground plane.
Such phased arrays have a long history, see e.g,~\cite{Allen1962}, and have recently been applied to above 20\,GHz 5G communication~\cite{Ojaroudiparchin2016,Dzagbletey+eta2018,Liao2019}. The Q-factor model is a good tool for estimating the impedance bandwidth of phased arrays, as it allows the formulation of tractable optimization problems.  Here we determine lower 
bounds on the Q-factor under various design constraints. The aim here is to examine how well the Q-factor bounds predict the bandwidth, and to investigate the impact of different kinds of design constraints. The unit-cell representation of the arrays is used.

The Q-factor approach was used to derive bounds for small antennas contained in a sphere or a cylinder in the works of Wheeler~\cite{Wheeler1947} and Chu~\cite{Chu1948} in the late 1940s.
The Q-factor bounds received a renewed interest, and many useful results in the last decade for small single-port antennas contained in arbitrarily shaped volumes~\cite{Vandenbosch2011,Tayli+etal2018,Tayli+Gustafsson2015,Gustafsson+tutorial2016,Capek+etal2019,Gustafsson+etal2012,Shi+etal2017,Gustafsson+Nordebo2013,Gustafsson+Capek+Schab2019}.
For small antennas, the Q-factor bounds have been used to construct trade-off relations between the Q-factor and  the  size~\cite{Capek+etal2017}, the  form-factor~\cite{Gustafsson+Nordebo2013}, the medium losses~\cite{Gustafsson+Capek+Schab2019}, and the radiation characteristics~\cite{Jonsson+etal2017}.

The Q-factor is defined to be proportional to a ratio of energies, stored and dissipated per cycle~\cite{IEEEstandard2014}, and can hence be represented in the electromagnetic  fields~\cite{Collardey+etal2005,Sten2004,Sten+etal2001,Fante1969,Collin+Rothschild1964}, in the current densities~\cite{Harrington+Mautz1972,Vandenbosch2010,Vandenbosch2013,Vandenbosch2013b,Geyi2003,Gustafsson+Ehrenborg2017,Capek+etal2016}, and in the system-level quantities~\cite{Smith1969,Polevoi1990,Chu1948}.
A suitable representation is the key for obtaining Q-factor bounds. 
The Chu bound~\cite{Chu1948} (see also~\cite{Harrington1960,Thal1978}) was obtained from a Q-factor of a circuit model for spherical wave expansion of the fields outside a spherical volume containing an antenna.
Collin and Rothschild~\cite{Collin+Rothschild1964} separated the total energy into stored and radiated by subtracting radiating spherical (or cylindrical) modes at the field level, which was extended by Fante to obtain Q-factor bounds~\cite{Fante1969}.
These results, although powerful and insightful, have a drawback - the majority of the antennas do not conform to the shapes (sphere or cylinder) used in these bounds. Such bounds depend only on the antenna's electrical size (the radius/length of antenna's circumscribed sphere/cylinder normalized to wavelength).
To overcome this disadvantage, the precise but implicit connection of the Q-factor to antenna geometry was proposed by Vandenbosch~\cite{Vandenbosch2010}, with the Q-factor expressed in terms of antenna currents.
This work was a generalization of small-antenna expressions by Geyi~\cite{Geyi2003}, and was further investigated by Gustafsson and Jonsson~\cite{Gustafsson+Jonsson2015stored,Jonsson+Gustafsson2015}. These antenna-current representations~\cite{Vandenbosch2010,Gustafsson+Jonsson2015stored,Jonsson+Gustafsson2015} were used to obtain the bounds for antennas in arbitrarily shaped volumes~\cite{Gustafsson+etal2012,Vandenbosch2011,Cismasu+Gustafsson2014,Tayli+etal2018,Shi+etal2017,Gustafsson+Nordebo2013}.

For periodic structures, Edelberg and Oliner~\cite{Edelberg+Oliner1960,Edelberg+Oliner1960b} investigated Q-factor expressions for arrays of slots and dipoles. A different approach was considered by 
 Tomasic and Steyskal~\cite{Tomasic+Steyskal2007,Tomasic+Steyskal2007b}. They proposed a lower bound for the Q-factor of a one-dimensional array of infinitely long cylinders in free space and above the ground plane.
Their result is based on the dominant cylindrical mode and they express the stored energies in terms of field densities in a unit cell (the fields of the propagating Floquet mode are excluded from stored energies).
This approach can be seen as a periodic-structures counterpart of Collin and Rothschild's method~\cite{Collin+Rothschild1964}.
Kwon and Pozar~\cite{Kwon+Pozar2014} derived results for arrays of dipoles in free space, above a ground plane, and above a grounded dielectric substrate.
These results provide good qualitative and quantitative analysis for Q-factor dependence on array parameters such as periodicity or scanning angle.
An assumption in~\cite{Kwon+Pozar2014} is that the current shape is known. To determine Q-factor bounds it is interesting to consider a larger range of geometries and to dispense with the known current assumption.

In this paper, we apply optimization methods to the previously derived current-density representations for the array Q-factor~\cite{LudvigOsipov+Jonsson2019,LudvigOsipov+Jonsson2019gp} to minimize the Q-factor over the current densities. These current densities are confined to an arbitrarily chosen geometry.
By imposing optimization constraints on the shape, medium losses, and polarization characteristics, we obtain the trade-off relations. 
The bounds for beam-scanning capabilities are presented.
We use two different optimization methods providing global minima to optimization problems.
One of the optimization approaches is based on semi-definite relaxation, and the other is based on a small-sized eigenvalue problem associated with critical points to the Lagrangian of the original problem.

As mentioned above, it is known that the Q-factor can be used to estimate the bandwidth both for single-port antennas and for arrays. In~\cite{LudvigOsipov+Jonsson2019,LudvigOsipov+Jonsson2019gp} it was shown that Q-factors equal to or greater than $5$ tend to predict the bandwidth of the tested array antennas. In this paper, we compare the Q-factor bounds with a sequence of numerically simulated array antennas with specified port positions. The comparison focuses on narrow-band antenna elements. 

The rest of this paper is organized as follows.
Section~\ref{sec:stored_energies} defines the array geometries, gives Q-factor expressions for arrays in terms of current density both with and without a ground plane, and provides a matrix form for stored energy kernels in numerical implementation.
In Section~\ref{sec:optimization_methods} two approaches for Q-factor optimization are proposed: semi-definite relaxation and the eigenvalue-type-problem based method.
Examples of physical bounds obtained numerically by the optimization methods are provided in Section~\ref{sec:numerical_examples}. The paper ends with conclusions in Section~\ref{sec:conclusions}.

\section{Stored energies and Q-factor}
\label{sec:stored_energies}
The Q-factor is proportional to the ratio between the stored energy and the total dissipated power~\cite{IEEEstandard2014}
\begin{equation}
Q = \frac{2\omega \max \{ \We,\Wm\} }{P_\mrm{tot}}.
\label{eq:Q_We_Wm}
\end{equation}
Here, $\omega$ is the angular frequency, $\We$ and $\Wm$ are electric and magnetic stored energies respectively.
In the case of radiating systems, the total power $P_\mrm{tot}$ is the sum of the radiated power $\Pr$ and conductive losses $\Pohm$.

The radiation efficiency is
\begin{equation}\label{eq:rad}
    \rade = \frac{\Pr}{\Pr + \Pohm} = \frac{1}{1+\delta},
\end{equation}
where $\delta = \Pohm/\Pr$ is the dissipation factor~\cite{Harrington1960}.
The radiation Q-factor is $Q_{\rm rad}=Q/\rade$.

A tractable Q-factor optimization problem can be obtained by expressing  both of the stored energies and the radiated and dissipated powers in terms of the current densities. This process  starts in Section \ref{sec:QJ} for array elements in free space, and for arrays with a ground plane in Section~\ref{sec:GP}.

\subsection{Free space case}\label{sec:QJ}

\begin{figure}
	\centering
	\includegraphics[scale=0.9]{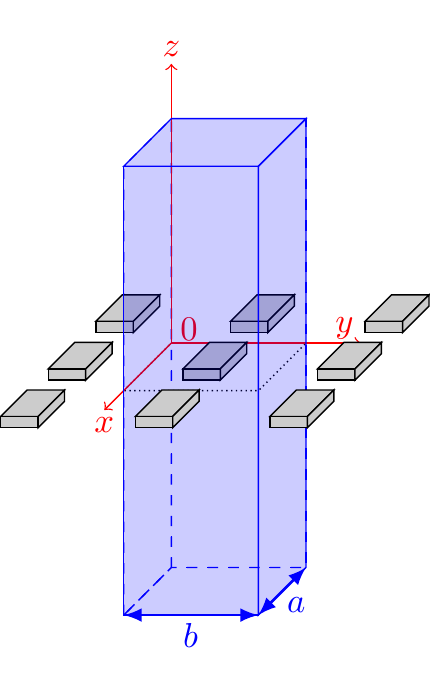}
\caption{An example of an array geometry for current density optimization. The optimization is performed over all possible electrical current densities on the surface of an array element (cuboid in this example) within a unit cell.}
\label{fig:array_sketch}
\end{figure}

Consider a three-dimensional array of perfectly electric-conducting (PEC)  elements on a two-dimensional rectangular grid, see Fig.~\ref{fig:array_sketch}. The PEC elements are of finite size, sufficiently regular and support an electric current density $\vec{J}(\vr)$ on their surface. The periodic solution satisfies the condition
\begin{equation}
\vec{J}(\vec{r}+\vec{\zeta}_{mn})=\vec{J}(\vec{r})\eu^{-\iu \Ktnull\cdot \vec{\zeta}_{mn}}. 
\label{eq:J_periodic}
\end{equation}
Here $\vec{r}\in\mathbb{R}^3$ is a coordinate vector, $\vec{\zeta}_{mn}=am\hat{\vec{x}}+bn\hat{\vec{y}}$; $m,n\in \Z$, and $\Ktnull=k\sin\theta_0\cos\phi_0\hat{\vec{x}}+k\sin\theta_0\sin\phi_0\hat{\vec{y}}$ is a phase shift vector with the polar $\theta_0$ and azimuthal $\phi_0$ angles and the wavenumber $k$.
The unit cell is defined as $U=\{(x,y,z)\in \R^3:x\in[0,a],y\in[0,b]\}$ with $a, b$ being the respective grid periods.

The stored energies in \eqref{eq:Q_We_Wm}, are typically defined in terms of the electric and the magnetic fields~\cite{Tomasic+Steyskal2007,Kwon+Pozar2014,LudvigOsipov+Jonsson2019,LudvigOsipov2020}. They are reformulated as quadratic forms of unit-cell's electric current density in~\cite{LudvigOsipov+Jonsson2019}, and rewritten here using dyadic-kernels notation.
For a PEC structure, as considered here, it serves to investigate the surface currents:
\begin{flalign}
    \We &=    \int_\Om \int_\Om 
\vec{J}^*(\vec{r}_1)\cdot{\rm \bf K}_{\rm e}(\vec{r}_1,\vec{r}_2)\cdot\vec{J}(\vec{r}_2)\diff S_1 \diff S_2,
\label{eq:We_Ke}\\
    \Wm &=    \int_\Om \int_\Om 
\vec{J}^*(\vec{r}_1)\cdot{\rm \bf K}_{\rm m}(\vec{r}_1,\vec{r}_2)\cdot\vec{J}(\vec{r}_2)\diff S_1 \diff S_2,
\label{eq:Wm_Km}
\end{flalign}
where $\Om$ is the maximal spatial support of current density in a unit cell, and the coordinates $\vr_1,\vr_2$ are integration variables corresponding to the surface elements $\diff S_1,\diff S_2$. The dyadic kernels are
\begin{flalign}
\begin{split}
    {\rm \bf K}_{\rm e}(\vec{r}_1,\vec{r}_2) =& 
    \frac{\mu}{4}\Re \left\{ \frac{-1}{k^2}\overline{\overline{\nabla}}_1\nabla_1 \G(\vec{r}_1,\vec{r}_2) + \Hd g(\vec{r}_1,\vec{r}_2)\right\},
    \label{eq:Ke_free_space}
\end{split}
\end{flalign}
\begin{flalign}
\begin{split}
    {\rm \bf K}_{\rm m}(\vec{r}_1,\vec{r}_2) =& 
    \frac{\mu}{4}\Re \left\{ \G(\vec{r}_1,\vec{r}_2)\unitdyadic + \Hd g(\vec{r}_1,\vec{r}_2)\right\},
    \label{eq:Km_free_space}
\end{split}
\end{flalign}
where, $\overline{\overline{\nabla}}_1$ is a Jacobi matrix,\footnote{
Let $G=G(\vr_1,\vr_2)$ then
$-\iint \vec{J}^*(\vec{r}_1)\cdot \overline{\overline{\nabla}}_1 (\nabla_1 G)\cdot \vJ(\vec{\vr_2}) \diff S_1\diff S_2 = \iint (\nabla_1 \cdot\vec{J}^*(\vec{r}_1))G(\nabla_2 \cdot \vec{J}(\vec{r}_2))\diff S_1\diff S_2$ for currents with support inside the unit-cell. 
}
$\nabla_{i}$ is a gradient with respect to $\vr_i,\; i=1,2$,
$\unitdyadic$ is a unit dyadic, $k$ is the wavenumber and $\mu$ is permeability of free space.
The free-space periodic Green's function $\G$ is given by~\eqref{eq:Greens_function} and the function $g$ is given by~\eqref{eq:little_g_exact}.
The dyadic-valued operator $\Hd = \left(  k^2\unitdyadic + \overline{\overline{\nabla}}_1\nabla_1 \right)$ is introduced for brevity.

Note here that both the Green's function and the function $g$ depend on the scan angle. Consequently, so do the stored energy kernels.

The radiated power in terms of the current densities is
\begin{equation}\label{eq:Pr}
    \Pr =   \frac{1}{2} \int_\Om \int_\Om 
\vec{J}^*(\vec{r}_1)\cdot{\rm \bf K}_{\rm r}(\vec{r}_1,\vec{r}_2)\cdot\vec{J}(\vec{r}_2)\diff S_1 \diff S_2,
\end{equation}
with the kernel
\begin{equation}
\begin{split}
    {\rm \bf K}_{\rm r}(\vec{r}_1,\vec{r}_2) = 
    \frac{\eta}{k}\Im \left\{ \Hd \G(\vec{r}_1,\vec{r}_2) \right\},
\end{split}
\end{equation}
 where $\eta$ is the free space impedance. 

Dissipation losses are here estimated by a perturbation approach \cite{Jackson1999,Gustafsson+Capek+Schab2019} with a surface resistance model
\begin{equation}
    \Pohm = 
    \frac{1}{2}\int_{\Om} R_{\rm s}(\vr_1)  |\vec{J}(\vr_1)|^2 \diff S_1,
    \label{eq:P_ohm}
\end{equation}
where $R_{\rm s}$ is a surface resistance.

\subsection{Ground plane case}\label{sec:GP}
The unit-cell geometry for the case with a ground-plane is similar to the unit-cell of the free-space case.
The array consists of three-dimensional PEC elements placed on a two-dimensional rectangular grid at positions $z>0$, with the infinite ground plane placed at $z=0$.
The unit cell is given by $U=\{(x,y,z)\in \R^3:x\in[0,a],y\in[0,b],z\in[0,\infty)\}$ and the phase shift condition~\eqref{eq:J_periodic} holds.
The stored energies for this configuration are derived in~\cite{LudvigOsipov+Jonsson2019gp} and written here using dyadic-kernel notation. 
The utilized mirroring approach results in  quadratic forms, see~\eqref{eq:We_Ke} and~\eqref{eq:Wm_Km} with updated $U$ and  where the kernel dyadics are given by 
\begin{equation}\label{KeM}
\begin{split}
&{\rm \bf K}_{\rm e}(\vec{r}_1,\vec{r}_2) = 
\frac{\mu}{4}\Re \left\{ \frac{-1}{k^2}\overline{\overline{\nabla}}_1\nabla_1 \G(\vec{r}_1,\vec{r}_2)  \right. \\
&+ \frac{1}{k^2}\overline{\overline{\nabla}}_1\nabla_1\G(\vec{r}_1,\vec{r}_{\rm 2i})\unitz+\Hd g(\vec{r}_1,\vec{r}_2)
-\left.   \left[\Hd g(\vec{r}_1,\vec{r}_{\rm 2i})\right]\unitz \right\},
\end{split}
\end{equation}
\begin{equation}\label{KmM}
\begin{split}
{\rm \bf K}_{\rm m}(\vec{r}_1,\vec{r}_2) =& 
\frac{\mu}{4}\Re \left\{ \G(\vec{r}_1,\vec{r}_2)\unitdyadic  - \G(\vec{r}_1,\vec{r}_{\rm 2i})\unitz \right. \\
&+
\Hd g(\vec{r}_1,\vec{r}_2)
-\left.  \left[\Hd g(\vec{r}_1,\vec{r}_{\rm 2i})\right]\unitz \right\}.
\end{split}
\end{equation} 
The extra terms in \eqref{KeM}-\eqref{KmM} as compared to the free-space case~\eqref{eq:Ke_free_space}-\eqref{eq:Km_free_space} represent the contribution due to the ground plane (given here in the form of the mirrored image terms).
The dyadic $\unitz = \vhat{x}\vhat{x} + \vhat{y}\vhat{y} - \vhat{z}\vhat{z}$ inverts the sign of $z$-component, and  $\vr_{\rm 2i} = \unitz\cdot\vr_2$ is an image coordinate.
The radiated power kernel is
\begin{equation}
\begin{split}
{\rm \bf K}_{\rm r}(\vec{r}_1,\vec{r}_2) =  
\frac{\eta}{k}\Im & \left\{ \Hd \G(\vec{r}_1,\vec{r}_2)-
[\Hd \G(\vec{r}_1,\vec{r}_{\rm 2i})]\unitz
 \right\}.
\end{split}
\end{equation}
The calculation of losses with the ground plane remains an open problem. 

\subsection{Polarization}

To incorporate the polarization-related constraints into the Q-factor optimization problem,
in this subsection we derive and formulate the polarization in terms of the current density.

Starting from the dyadic Green's function representation, the electric field due to a surface electric current density $\vec{J}$ is 
\begin{equation}
\begin{split}
\vec{E}(\vec{r}_1) = 
\frac{\iu \eta}{k}\int_{\Om}\left\{ -\Hd \G(\vec{r}_1,\vec{r}_2) \right\}\cdot \vec{J}(\vr_2) \diff S_2.
\end{split}
\label{eq:E_G_J}
\end{equation}
The contribution, associated with $(m,n)$:th Floquet mode of the Green's function~\eqref{eq:Greens_function}, is
\begin{equation}
\begin{split}
&\vec{E}_{mn}(\vec{r}_1) = \\
&\frac{ -\eta}{2 k ab}\int_{\Om}\left\{\Hd \frac{\eu^{-\iu\Ktmn\cdot(\vec{\rho}_1-\vec{\rho}_2)}\eu^{-\iu \Kzmn|z_1-z_2|}}{\Kzmn} \right\}\cdot \vec{J}(\vr_2) \diff S_2.
\end{split}
\label{eq:Emn}
\end{equation}
Here, the wave-vector and the coordinates are separated in the longitudinal and transverse components, see Appendix~\ref{sec:appendix}.
Calculating the expression within $\{\cdot \}$ in the integrand~\eqref{eq:Emn} reduces the mode to
\begin{equation}
    \vec{E}_{mn}(\vec{r}_1) = 
\frac{ \eta}{2 S}
\eu^{-\iu\Ktmn\cdot \vec{\rho}_1}\eu^{\mp\iu \Kzmn z_1}
\vec{F}_{mn\pm}.
\end{equation}
Here, $\vec{F}_{mn\pm}$ is proportional to the polarization of the $(m,n)$:th Floquet mode. The $+$ ($-$)-sign applies to 
 observation points $\vr_1$ above (below) the array, \ie such that $z_1 \geq z_2$ ($z_1\leq z_2$) for all $\vr_2\in \Om$. These factors have the form
\begin{equation}
    \vec{F}_{mn\pm} = \overline{\overline{\mathbf{K}}}_\pm\cdot 
        \int_{\Om}
        \eu^{\iu\Ktmn\cdot\vec{\rho}_2}\eu^{\pm\iu \Kzmn z_2}
        \vec{J}(\vr_2) \diff S_2.
        \label{eq:Fmn}
\end{equation}
The dimensionless dyadics $\overline{\overline{\mathbf{K}}}_\pm$ are given by
\begin{equation}
\overline{\overline{\mathbf{K}}}_\pm =
	 \frac{1}{k\Kzmn}
	\begin{bmatrix}
	\Kxm^2 - k^2 & \Kxm\Kyn & \pm\Kxm \Kzmn  \\
	\Kxm\Kyn & \Kyn^2 - k^2 & \pm\Kyn \Kzmn   \\
	 \pm\Kxm \Kzmn  & \pm\Kyn \Kzmn  & \Kzmn^2 - k^2 \\
	\end{bmatrix} .
	\label{eq:K_plus}
\end{equation}

Recall that the wave-vector components $\Ktmn=\vhat{x} \Kxm+\vhat{y}\Kyn,\Kzmn$ depend on the angles $\phi_0,\theta_0$ associated  with the direction of propagation of the Floquet modes. Thus, $\overline{\overline{\mathbf{K}}}_\pm$ depend both on the electrical sizes $ka$, $kb$ of the unit cell and the angles $\phi_0,\theta_0$ associated with the propagation direction of the investigated Floquet mode, as specified in $\Ktnull$, see the Appendix.

The co- and cross-polarized components of a Floquet mode in a given direction are (analogous to the far-field definition~\cite{Ludwig1973})
\begin{equation}
    F_{{\rm co},mn\pm}=
    \vec{F}_{mn\pm}\cdot \vhat{e}^*_{\rm co},\ \ \ F_{{\rm cx},mn\pm}=
    \vec{F}_{mn\pm}\cdot \vhat{e}^*_{\rm cx},
    \label{eq:co+crosspol}
\end{equation}
where the reference polarizations $\vhat{e}_{\rm co}$, and $\vhat{e}_{\rm cx}$ are the chosen co- and cross-polarization unit vectors.

For the ground-plane case, there is no radiation below the array, that is $\vec{F}_{mn-}=0$. 
The factor $\vec{F}_{mn+}$ of the Floquet-modes above the array with ground plane is
\begin{multline}
    \vec{F}_{mn+} = \overline{\overline{\mathbf{K}}}_+\cdot 
        \int_{\Om}
        \eu^{\iu\Ktmn\cdot\vec{\rho}_2}\eu^{\iu \Kzmn z_2}
        \vec{J}(\vr_2) \diff S_2
        \\ -
        \overline{\overline{\mathbf{K}}}_{\rm im}\cdot 
        \int_{\Om}
        \eu^{\iu\Ktmn\cdot\vec{\rho}_2}\eu^{-\iu \Kzmn z_2}
        \vec{J}(\vr_2) \diff S_2.
        \label{eq:Fmn_gp}
\end{multline}
Here, $\overline{\overline{\mathbf{K}}}_+$ is given in~\eqref{eq:K_plus} and
\begin{equation}
\overline{\overline{\mathbf{K}}}_{\rm im} =
	 \frac{1}{k\Kzmn}
	\begin{bmatrix}
	\Kxm^2 - k^2 & \Kxm\Kyn & -\Kxm \Kzmn  \\
	\Kxm\Kyn & \Kyn^2 - k^2 & -\Kyn \Kzmn   \\
	 \Kxm \Kzmn  & \Kyn \Kzmn  & k^2 - \Kzmn^2 \\
	\end{bmatrix} .
\end{equation}
The co-polarized and cross-polarized components are found similarly to the free-space case from~\eqref{eq:co+crosspol}.

\subsection{Matrix formulation}
\label{sec:matrix_formulation}

To efficiently determine these electromagnetic quantities numerically, all the above linear and quadratic forms are reduced to finite-dimensional linear and quadratic forms with matrix kernels.
To this end, the current density is approximated by a set of surface basis functions $\{\vec{f}_u(\vec{r})\}_{u=1}^N$ over the array element $\Omega$
\begin{equation}
    \vec{J}(\vec{r}) \approx \sum_{u=1}^{N} I_u \vec{f}_u(\vec{r}).
\end{equation}
Here, the Rao-Wilton-Glisson basis functions are used for $\{\vec{f}_u(\vec{r})\}_{u=1}^N$. The stored-energies and radiated-power kernels, in both free-space and ground-plane cases, can be represented as matrices with elements (the upper index $u,w\in [1,N]$ below corresponds to element indices in the corresponding matrix)
\begin{equation}
\begin{Bmatrix} 
{\rm \bf W}_{\rm e}^{(u,w)}  \\ 
{\rm \bf W}_{\rm m}^{(u,w)}  \\ 
\Rbf^{(u,w)}  
\end{Bmatrix}
=\int_\Om \int_\Om 
\vec{f}_u(\vec{r}_1)\cdot
\begin{Bmatrix}
{\rm \bf K}_{\rm e}(\vec{r}_1,\vec{r}_2) \\
{\rm \bf K}_{\rm m}(\vec{r}_1,\vec{r}_2) \\
{\rm \bf K}_{\rm r}(\vec{r}_1,\vec{r}_2)
\end{Bmatrix}
\cdot\vec{f}_w(\vec{r}_2)\diff S_1 \diff S_2.
\end{equation}
The stored energies \eqref{eq:We_Ke}-\eqref{eq:Wm_Km} and radiated power \eqref{eq:Pr} are thus approximated as
\begin{equation}
    \We \approx {\rm {\bf I}^H {\bf W}_e {\bf I}}, \quad 
    \Wm \approx {\rm {\bf I}^H {\bf W}_m {\bf I}}, \quad 
    \Pr \approx \frac{1}{2}{\rm {\bf I}^H \Rbf {\bf I}},
    \label{eq:Matrices}
\end{equation}
with the current density basis coefficients vector ${\rm \bf I}=(I_1, I_2, ..., I_N)^{\rm T}$, and $\{.\}^{\rm T}$ and $\{.\}^{\rm H}$ denote a transpose and a conjugate transpose respectively.
The process of computing the stored-energies and radiated-power matrices has a similar structure to the impedance-matrix computation in the Method of Moments (MoM) and can be implemented by marginal modifications of a MoM code~\cite{LudvigOsipov+Jonsson2019,LudvigOsipov+Jonsson2019gp}.
The Ohmic losses \eqref{eq:P_ohm} are represented as a quadratic form $\Pohm = \frac{R_{\rm s}}{2}{\rm {\bf I}^H {\bf \Psi} {\bf I}}$, where ${\bf \Psi}$ is the Gram matrix~\cite{Gustafsson+Capek+Schab2019}. The Gram matrix has elements
\begin{equation}
{\bf \Psi}^{(u,w)} = \int_\Omega \vec{f}_w(\vec{r}_1)\cdot\vec{f}_w(\vec{r}_1)\diff S_1 
\label{eq:Gram}
\end{equation}

The polarizations of the Floquet-modes~\eqref{eq:Fmn} and~\eqref{eq:Fmn_gp} are vector-valued linear forms of the surface current density $\vec{J}$.
The projections~\eqref{eq:co+crosspol} of the Floquet-modes on the co- and cross-polarized unit vectors are approximated as
\begin{equation}
     F_{{\rm co},mn\pm} \approx {\bf F}_{{\rm co},mn\pm} {\bf I},
\qquad
     F_{{\rm cx},mn\pm} \approx {\bf F}_{{\rm cx},mn\pm} {\bf I}.
     \label{eq:Fvectors}
\end{equation}
The elements of vectors ${\bf F}_{{\rm co},mn\pm}$ and ${\bf F}_{{\rm cx},mn\pm}$ are given by (here demonstrated for co- and cross-polarized waves in the free space) 
\begin{equation}
\begin{Bmatrix} 
\mathbf{F}_{{\rm co},mn\pm}^{(u)}  \\
\mathbf{F}_{{\rm cx},mn\pm}^{(u)}   
\end{Bmatrix}
=
\begin{Bmatrix} 
\vhat{e}_{\rm co}^*  \\ 
\vhat{e}_{\rm cx}^*   
\end{Bmatrix}
\cdot
\overline{\overline{\mathbf{K}}}_\pm\cdot 
\int_\Om 
\eu^{\iu\Ktmn\cdot\vec{\rho}_2}\eu^{\pm\iu \Kzmn z_2}
\vec{f}_u(\vec{r}_2) \diff S_2.
\end{equation}

\section{Optimization methods}
\label{sec:optimization_methods}

\subsection{Q-factor optimization}

In the previous section, the operator representation of the stored energies and radiated power were reduced to finite-sized matrices. This reduction simplifies the Q-factor representation \eqref{eq:Q_We_Wm} to a ratio between two finite-dimensional quadratic forms. A Q-factor optimization  problem for a given shape and scanning direction can be formulated as:
\begin{equation}
	 Q_*=\minimize_{{\bf I}\in \mathbb{C}^N}\ Q({\bf I}) ,
\label{Qratio}
\end{equation}
where $Q({\bf I})$ is the Q-factor, expressed in terms of the current coefficients. The optimization is over all currents that can be represented by the coefficients $I_n$ in {\bf I}.
The above problem is scaling invariant under the transformation ${\bf I}\mapsto a{\bf I}$, $a\in \C\setminus\{0\}$ \cite{Gustafsson+etal2012}. Thus, the amplitude of the current and its overall phase does not change the Q-factor. 
This implies that a degree of freedom in the optimization problem can be removed: it suffices to investigate all currents for which the unit-cell radiates half a watt.
Consequently, this allows~\eqref{Qratio} to be written as
\begin{equation}
\begin{aligned}
	Q_* = & \minimize_{{\bf {I}}\in \mathbb{C}^N} && Q \\
	& \subto &&  4\omega {\rm {\bf I}^H {\bf W}_e {\bf I}} \leq Q, \\
	&  &&  4\omega {\rm {\bf I}^H {\bf W}_m {\bf I}} \leq Q, \\
	& && {\rm {\bf I}^H \Rbf {\bf I}} = 1.
\end{aligned}
\label{Qquad}
\end{equation}
 The above Q-factor optimization gives the minimum Q-factor at a preferred scan-direction and enclosing shape, as determined by the support  of the set of basis functions.

Additional design constraints can now be added to~\eqref{Qquad}.  
For example, the constraint on efficiency
\begin{equation}
    \frac{R_{\rm s}}{2}{\rm {\bf I}^H {\bf \Psi} {\bf I}} = \delta,
    \label{eq:P_ohm_constr}
\end{equation}
can be added to~\eqref{Qquad}.
Here $\delta = \Pohm/\Pr$ is the dissipation factor, see the discussion below~\eqref{eq:rad}.
Similarly, a constraint on the polarization purity for a specified direction $\phi_0,\theta_0$ can be formulated as:
\begin{equation}
    |{\bf F}_{{\rm cx},mn\pm}\Ibf| \leq \xi |{\bf F}_{{\rm co},mn\pm}\Ibf|.
\end{equation}
This can in turn be reformulated as a quadratic constraint
\begin{equation}
\Ibf^\herm{\bf F}_{{\rm cx},mn,\pm}^\herm {\bf F}_{{\rm cx},mn,\pm}\Ibf \leq 
    \xi^2 \Ibf^\herm{\bf F}_{{\rm co},mn,\pm}^\herm {\bf F}_{{\rm co},mn,\pm}\Ibf.
    \label{eq:F_constr}
\end{equation}

The above problem~\eqref{Qquad} with (and without) the optional constraints~\eqref{eq:P_ohm_constr} and/or~\eqref{eq:F_constr} is a quadratically constrained quadratic program (QCQP), which is a well-studied class of optimization problems. In the context of Q-factor optimization, see~\cite{Jonsson+etal2017,Shi2018Thesis,Gustafsson+Ehrenborg2017}. 

\subsection{Semi-definite relaxation}

One method to solve QCQP is the semi-definite relaxation (SDR) \cite{Goemans+Williamson1995}. 
It is based on the identity
\begin{equation}\label{tr}
    \Ibf^\herm \Abf \Ibf = \trace (\Ibf^\herm \Abf \Ibf) = \trace (\Abf \Ibf \Ibf^\herm) = \trace (\Abf \Xbf),
\end{equation}
where $\Xbf =\Ibf \Ibf^\herm$ and $\trace(\cdot)$ denotes the trace.
The Q-factor minimization problem \eqref{Qquad} is then with help of the identity~\eqref{tr} reformulated as a semidefinite relaxation program
\begin{equation}
\begin{aligned}
Q_* =	& \minimize_{0\preccurlyeq\Xbf\in\mathcal{H}} && Q \\
	& \subto &&  4\omega\trace(\Webf\Xbf) \leq Q, \\
	& &&         4\omega\trace(\Wmbf\Xbf) \leq Q, \\
	& &&         \trace(\Rbf\Xbf) =1,
\end{aligned}
\label{eq:SDR}
\end{equation}
where the condition $\rank{\Xbf}=1$ is dropped, and $\mathcal{H}$ denotes the class of $N\times N$ Hermitian matrices.
Whenever the solution has $\rank (\Xbf)=1$, one can reconstruct~\cite{Luo+etal2010} the current density $\Ibf$ from the non-zero eigenvalue $\lambda$ and the corresponding eigenvector $\rm{\bf{u}}$ of $\Xbf$. The corresponding optimizing current-density-coefficients vector is obtained as $\Ibf_* = \lambda^{1/2}{\rm\bf u}$.

The constraints~\eqref{eq:P_ohm_constr} and~\eqref{eq:F_constr} are formulated as
\begin{equation}
    R_{\rm s}\trace({\bf \Psi}\Xbf) = \delta,
\end{equation}
\begin{equation}
\trace ({\bf F}_{{\rm cx},mn,\pm}^\herm {\bf F}_{{\rm cx},mn,\pm} \Xbf - \xi^2{\bf F}_{{\rm co},mn,\pm}^\herm {\bf F}_{{\rm co},mn,\pm} \Xbf) \leq 0.
\end{equation}

\subsection{The eigenvalue method}

An alternative and less memory demanding approach is based on a Lagrangian  formulation of the optimization problem. An alternative form of~\eqref{Qquad} is
\begin{equation}\label{eq:Qmax}
\begin{aligned}
\frac{Q_*}{4\omega}=	& \minimize_{\Ibf\in\C^N} && \max(\Ibf^\herm\Webf \Ibf, \Ibf^\herm\Wmbf \Ibf), \\
	& \subto &&    \Ibf^\herm\Rbf \Ibf =1.
\end{aligned}
\end{equation}
The maximum of the stored energies is an upper bound for the convex combination ($\alpha \in [0;1]$) of the electric and magnetic stored energies~\cite{Gustafsson+tutorial2016}
\begin{equation}\label{Wa}
    \max(\Ibf^\herm\Webf \Ibf, \Ibf^\herm\Wmbf \Ibf) \geq \Ibf^\herm\Walpha \Ibf,
\end{equation}
where
\begin{equation}
    \Walpha  = \alpha \Webf + (1-\alpha) \Wmbf, \ \alpha\in[0,1].
\end{equation}
The minimization problem~\eqref{eq:Qmax} is then  relaxed to
\begin{equation}
\begin{aligned}
	\frac{Q_{\rm R}}{4\omega}=& \max_\alpha \min_\Ibf && \Ibf^\herm\Walpha \Ibf \\
	& \subto &&    \Ibf^\herm\Rbf \Ibf =1.
\end{aligned}
\label{eq:maxmin}
\end{equation}
The Lagrangian functional associated with this problem is
\begin{equation}\label{eq:Lag}
    L(\alpha,\lambda) = \Ibf^\herm\Walpha \Ibf + \lambda (1 - \Ibf^\herm\Rbf \Ibf).
\end{equation}
Critical points of the Lagrangian are found as zeros of the variational derivative
of the Lagrangian, i.e., the solutions to equation
\begin{equation}\label{eq:EV}
    \Walpha \Ibf  - \lambda\Rbf \Ibf = 0.
\end{equation}
Such critical points provide candidates to the solution of the dual problem. Here, the minimum generalized eigenvalue and eigenvector solution $(\lambda_*,\Ibf_*)$  minimizes the ratio $\Ibf^\herm\Walpha \Ibf/\Ibf^\herm\Rbf \Ibf$, which with the help of~\eqref{Wa} provides a lower bound to the Q-factor. Maximizing over the parameter $\alpha$ give us the maximal lower bound $Q_{\rm R}\leq Q_*$.

A straightforward solution of the resulting generalized eigenvalue problem \eqref{eq:EV} is usually ill-conditioned, due to the low rank of $\Rbf$, corresponding to the number of the propagating Floquet modes. This is mitigated by an eigendecomposition of the radiated power matrix $\Rbf={\rm \bf U}\tilde{\Rbf}{\rm \bf U}^\herm$, where ${\rm \bf U}$ is a unitary matrix, composed of the eigenvectors of $\Rbf$, and $\tilde{\Rbf}$ is a diagonal matrix with few non-zero entries due to the low rank of $\Rbf$:
\begin{equation}
    \tilde{\Rbf} = {\rm \bf U}^\herm \Rbf {\rm \bf U} = 
    \diag(d_1, ... , d_r, 0, ..., 0) =
  \begin{bmatrix}
    \tilde{\Rbf}_{11} & {\bf 0} \\
    {\bf 0} & {\bf 0}  
  \end{bmatrix}.
\end{equation}
Here, $r=\rank \Rbf$.
The currents can be represented as
\begin{equation}
    \Ibf = {\rm \bf U} \Tilde{\Ibf}.
    \label{eq:ItildeI}
\end{equation}
Multiplication of the eigenvalue problem~\eqref{eq:EV} by ${\rm \bf U}^\herm$ from the left gives
\begin{equation}
   {\rm \bf U}^\herm \Walpha {\rm \bf U} \Tilde{\Ibf} = \lambda {\rm \bf U}^\herm\Rbf {\rm \bf U} \Tilde{\Ibf}.
\end{equation}
The transformed stored energy matrix is then partitioned according to the eigendecomposition $\tilde{\Rbf}$ of the radiated power matrix:
\begin{equation}
    \tilde{\bf W}_\alpha = {\rm \bf U}^\herm \Walpha {\rm \bf U} = 
  \begin{bmatrix}
    \tilde{\Wbf}_{11} & \tilde{\Wbf}_{12} \\
    \tilde{\Wbf}_{21} & \tilde{\Wbf}_{22}  \\
  \end{bmatrix}.
\end{equation}
The generalized eigenvalue problem is then rewritten as a system of equations 
\begin{equation}\label{WI}
    \systeme{\tilde{\Wbf}_{11} \tilde{\Ibf}_1 + \tilde{\Wbf}_{12} \tilde{\Ibf}_2 =
    \lambda \tilde{\Rbf}_{11} \tilde{\Ibf}_1, 
    \tilde{\Wbf}_{21} \tilde{\Ibf}_1 + \tilde{\Wbf}_{22} \tilde{\Ibf}_2 = 0 }
\end{equation}
The stored energies are positive definite in the considered cases, thus the second equation provides the relation between the parts of $\tilde{\Ibf}$
\begin{equation}
    \tilde{\Ibf}_2 = - \tilde{\Wbf}_{22}^{-1} \tilde{\Wbf}_{21} \tilde{\Ibf}_1.
    \label{eq:Itilde2}
\end{equation}
Direct substitution in the upper equation of \eqref{WI} gives an eigenvalue problem of the size of $\tilde{\Rbf}_{11}$ which is $r\times r$:
\begin{equation}\label{rev}
    (\tilde{\Wbf}_{11} - \tilde{\Wbf}_{12}\tilde{\Wbf}_{22}^{-1} \tilde{\Wbf}_{21}) \tilde{\Ibf}_1  =
    \lambda \tilde{\Rbf}_{11} \tilde{\Ibf}_1.
\end{equation}
The lowest eigenvalue and its corresponding eigenvector in \eqref{rev} are denoted as  $(\lambda_*,\tilde{\Ibf}_{1*})$.
The best solution current $\Ibf_*$ to~\eqref{eq:EV} is found from $\tilde{\Ibf}_{1*}$ via~\eqref{eq:Itilde2},~\eqref{eq:ItildeI}.
After re-scaling to comply with the radiated-power constraint in~\eqref{eq:maxmin}, it solves the minimization problem in~\eqref{eq:maxmin} for a given $\alpha$.  The reduction~\eqref{eq:Itilde2} is essential to improve the accuracy of the solution $(\lambda_*,\tilde{\Ibf}_{1*})$.

Additional constraints in the eigenvalue approach are included by adding the constraints with their associated Lagrange multiplier to the Lagrangian \eqref{eq:Lag}. The eigenvalue approach tend to be faster than the SDR, since it has fewer unknowns as compared to \eqref{eq:SDR}. It however comes with an overhead of maximizing with respect to $\alpha$ and resolving possible degenerate of eigen-currents.
In our investigated cases, we find that the SDR-method and the eigenvalue method give the same minimum-value, indicating that these problems do not have a duality gap, i.e. $Q_*=Q_{\rm R}$. This follows since the SDR-method finds the true minimum with three or fewer constraints~\cite{Huang+Zhang2007}. Both methods have been applied in  Section~\ref{sec:numerical_examples}.


\section{Numerical examples}
\label{sec:numerical_examples}

Above, all the theory required to obtain the Q-bounds from current density optimization has been discussed. To investigate the Q-factor bounds, the following shapes and scenarios are considered: cuboids and planar rectangular structures of horizontal and vertical orientation and for different aspect ratios with and without a ground plane. All these examples have a square-sized unit-cell with side $p$, and the array elements are relatively small to focus on the narrow-band regime. Both the broadside case, as well as scanning are discussed. The impact of efficiency and different demands of polarization purity are also considered below.

In all the examples, except where otherwise stated, the numerical approximations of the stored energies, radiated power, the Gram matrix and the polarization vectors are computed as described in Section~\ref{sec:matrix_formulation} using RWG basis functions on a triangular mesh, representing considered array element geometries.
These numerical approximations are then used in conjunction with the optimization methods described in Section~\ref{sec:optimization_methods}
to obtain the Q-factor bounds with different constraints.

\subsection{Rectangular plates in free space}\label{ssec:Free}

\begin{figure}
	\centering
	\includegraphics{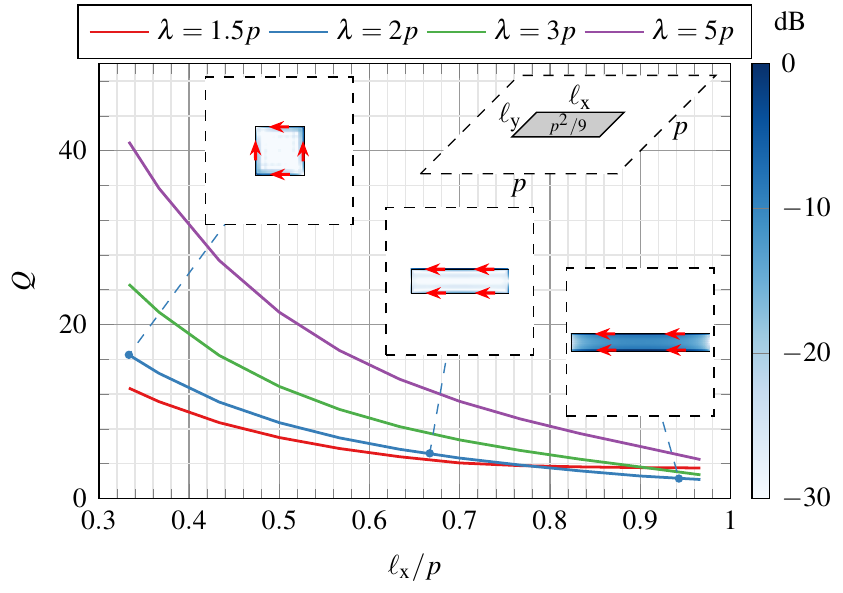}
\caption{The lower bound on the Q-factor for a sequence of rectangular plates with varied aspect ratio, constant total plate area $S=p^2/9$, period $p$ and wavelengths $\lambda$.
The optimal normalized current densities are in the insets, with the color bar showing the amplitude in dB.}%
\label{fig:Lx_Ly}
\end{figure}

First, consider a sequence of flat periodically spaced rectangles, $\ell_{\rm x}\times \ell_{\rm y}$, with constant area: $\ell_{\rm x}\ell_{\rm y}=p^2/9$, see Fig.~\ref{fig:Lx_Ly}. Here, $p$ is the length of unit-cell in both the $x$- and the $y$-direction, i.e.\ $p = a = b$.
The minimum Q-factor at broadside-scan for an array of rectangular PEC plates is depicted in Fig.~\ref{fig:Lx_Ly} as a function of the normalized length of the plate, $\ell_{\rm x}/p$.

Observe in Fig.~\ref{fig:Lx_Ly} that for this fixed small area result, it is clear that a larger length $\ell_{\rm x}$ gives a lower Q-factor. That is, the best-Q-factor design, enclosed in a square surface has a larger Q-factor than the best design enclosed in an oblong rectangular surface. Hence, to design a planar array with larger bandwidth (\ie lower Q-factor) with a limited surface area of the element, one should look into oblong element designs, rather than the ones that are more fitted for the square shape. Consequently, this narrow-band-design result with a fixed small area is different from the large wide-band ones where self-complementary designs are common. 
It is also interesting to note that optimal current solutions have their highest amplitude along the edge of the rectangle (see the inset in Fig.~\ref{fig:Lx_Ly}), and as the rectangle becomes more narrow, the whole structure is utilized. 

\subsection{Horizontal plate over a ground plane}

\begin{figure}
	\centering
	\includegraphics[width=\columnwidth]{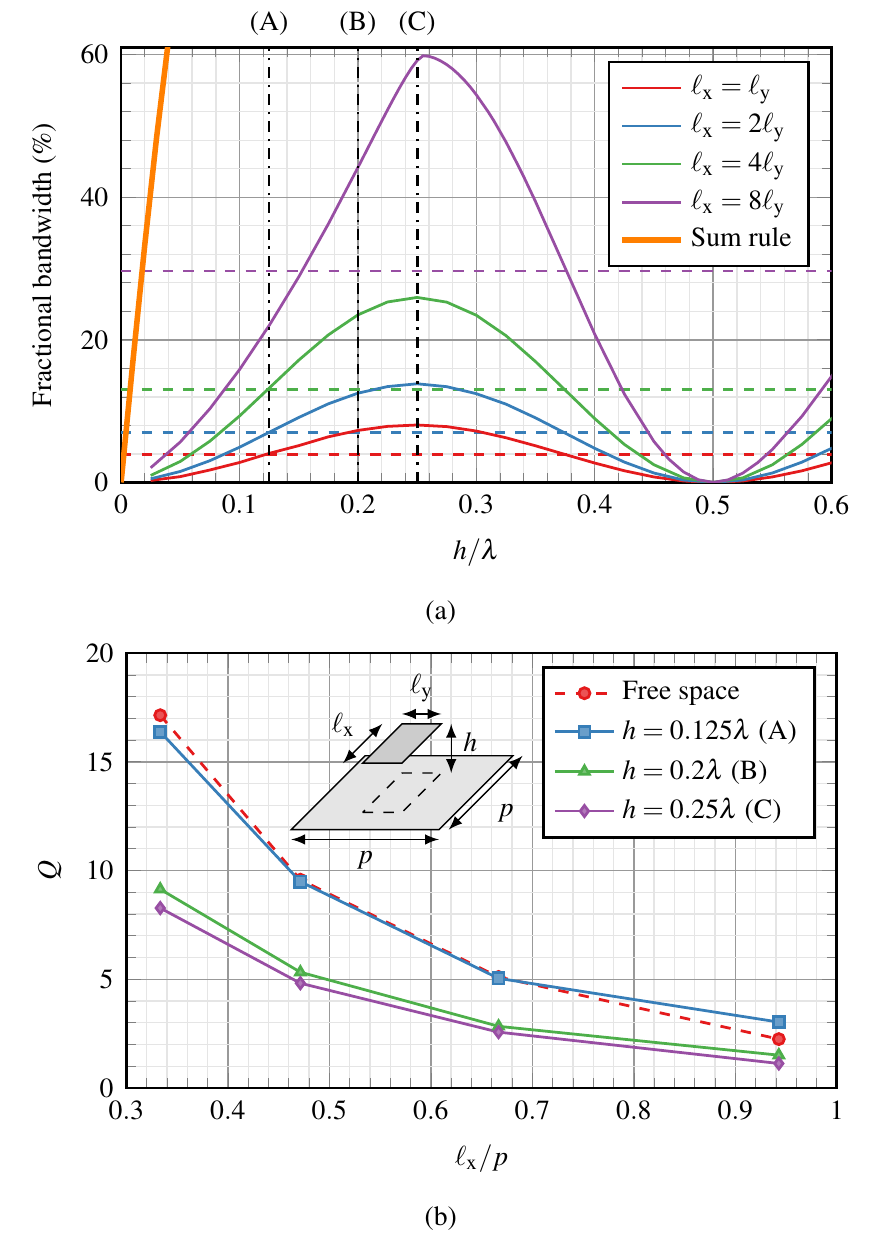}
\caption{(a) Q-factor-predicted $-$10-dB fractional bandwidth for rectangular plates of varying aspect ratios over a ground plane when $\lambda = 2p$ and $\ell_{\rm x} \ell_{\rm y}=p^2/9$. (b) The corresponding Q-factors as a function of $\ell_{\rm x}$ for heights marked in (a) with (A), (B), and (C).}
\label{fig:Lx_Ly_height}
\end{figure}

Next, a ground plane is added at $z=0$ to the structure in Sec.~\ref{ssec:Free}, see the inset in Fig.~\ref{fig:Lx_Ly_height}(b). This is an interesting case, since a Q-based bandwidth estimates can be compared with the array figure of merit for lossless antennas. It is known that the array figure of merit tends to well predict the available bandwidth for wide-band arrays~\cite{Jonsson+etal2013,Doane+etal2013}. The figure-of-merit result is based on a sum-rule~\cite{Nedic+etal2020} for planar arrays over a ground plane.

In all investigated cases, the area $p^2/9$ is fixed, and the wavelength at the center frequency is $\lambda=2p$. To convert the Q-factor to bandwidth, the Yaghjian and Best relation \cite{Yaghjian+Best2005} is used with a $-10$\,dB reflection coefficient. The Q-factor-estimated bandwidth
for different aspect ratios of rectangles, as a function of the distance from the ground plane, is depicted in Fig.~\ref{fig:Lx_Ly_height}(a). The dashed horizontal lines of the corresponding color show the estimated bandwidth for the equivalent rectangle in free space (no ground plane).
Depending on the distance from the ground plane, the ground plane can either enhance or degrade the bandwidth as compared to the free-space case. The widest Q-factor bandwidth estimate is obtained at $h = \frac{1}{4}\lambda$ for all aspect ratios of the fixed area.
This is consistent with the common rule of thumb of locating the antenna at a quarter of a wavelength above the ground plane.

The bandwidth for $\ell_{\rm x}\in [1,2,4]\ell_{\rm y}$ is equal to the free space case when the height is $h = \frac{1}{8}\lambda$ and $h = \frac{3}{8} \lambda$.
Below $h = \frac{1}{8}\lambda$ and above $h = \frac{3}{8} \lambda$ the ground plane reduces the available bandwidth. The more wide-band case with $\ell_{\rm x}=8\ell_{\rm 8}$ is here predicted to be better than the free-space case for a slightly smaller region, at the same time observe here, that the Q-factor is rather small.
At $h = \frac{1}{2}\lambda$ the Q-factor predicted bandwidth becomes essentially zero.

Comparing the Fig.~\ref{fig:Lx_Ly_height}(b) with the result in Fig.~\ref{fig:Lx_Ly} shows that the results for the plate above a ground plane follow the same trend as for the free space. The larger the ratio of the two sides of the rectangle, the lower Q-factor is obtained.
This is the case for all distances above the ground plane.

From a comparison of the Q-factor predicted bandwidth with the array figure of merit (Sum rule in Fig.~\ref{fig:Lx_Ly_height}(a)), it is clear that the Q-factor predicts a smaller bandwidth. The sum-rule, as used here, does not account for a particular shape of the element but only the volume of the array above the ground-plane. Initially, the sum-rule based bound grows linearly with the height $h$ illustrating that a good utilization of the unit-cell volume can have a wide bandwidth. 

\begin{figure}
\centering
	\includegraphics[width=\columnwidth]{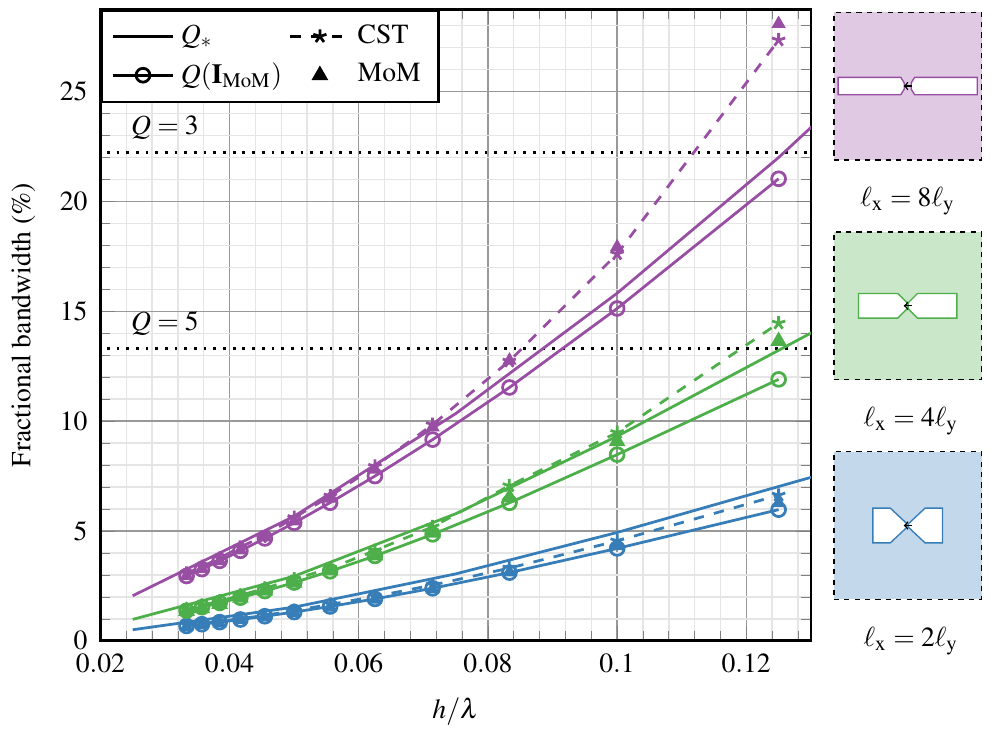}
	\caption{An enlargement of Fig.~\ref{fig:Lx_Ly_height}(a) with a comparison between bandwidth prediction and realized bandwidth. The dipoles to the right are the distance $h$ above the ground-plane. The color of the designs correspond to the curves colors.%
	}%
\label{fig:comparison}
\end{figure}

With focus on narrow-band arrays, it is therefore interesting to examine how certain realized resonant antenna elements behave as compared with the Q-factor and sum-rule bounds. A partial enlargement of Fig.~\ref{fig:Lx_Ly_height}(a) with the simulation results is shown in Fig.~\ref{fig:comparison}. 

Here, three dipole-like antenna elements with different aspect ratios, see Fig.~\ref{fig:comparison}, have been simulated both with the CST frequency domain solver and an in-house MoM-code. The obtained  impedance is then tuned to resonance with a single reactive lumped element and the $-10$\,dB bandwidth is determined. 
The single element tuned bandwidth\footnote{More advanced impedance-tuning naturally results in a wider bandwidth, and such improvement-factors are similar for these elements.} is a measure of antenna bandwidth that can be compared with the Q-factor predicted bandwidth~\cite{Yaghjian+Best2005}. 
The tuned CST- and MoM-bandwidths are depicted in Fig.~\ref{fig:comparison} with the $\star$-marked dashed lines and by $\blacktriangle$-markers respectively. These different software results agree well. 
The MoM-calculated currents of these gap-feed antennas are used to calculate a Q-factor by a substitution of the current into~\eqref{eq:Q_We_Wm}. The resulting bandwidth is depicted in Fig.~\ref{fig:comparison} with circle-marked solid lines. Note that they, as expected from the optimization problem formulation, are slightly below the bandwidth from optimal Q-factor (solid lines).

 Fig.~\ref{fig:comparison} shows that the Q-factor bounds here tend to well predict the obtained bandwidth. As the element gets more wide-band, e.g., $Q\lesssim 5$, the bandwidth predicted by the single-resonant-Q-factor model gradually starts to underestimate the available bandwidth.
These results remain well below the sum-rule based bound, as expected with the small utilization of the unit-cell area.
The Q-factor bounds presented here thus provide a better prediction for small {\it resonant} type array antenna elements.

\subsection{Ground plane with additional elements.}

\begin{figure}
\centering
\includegraphics[width=\columnwidth]{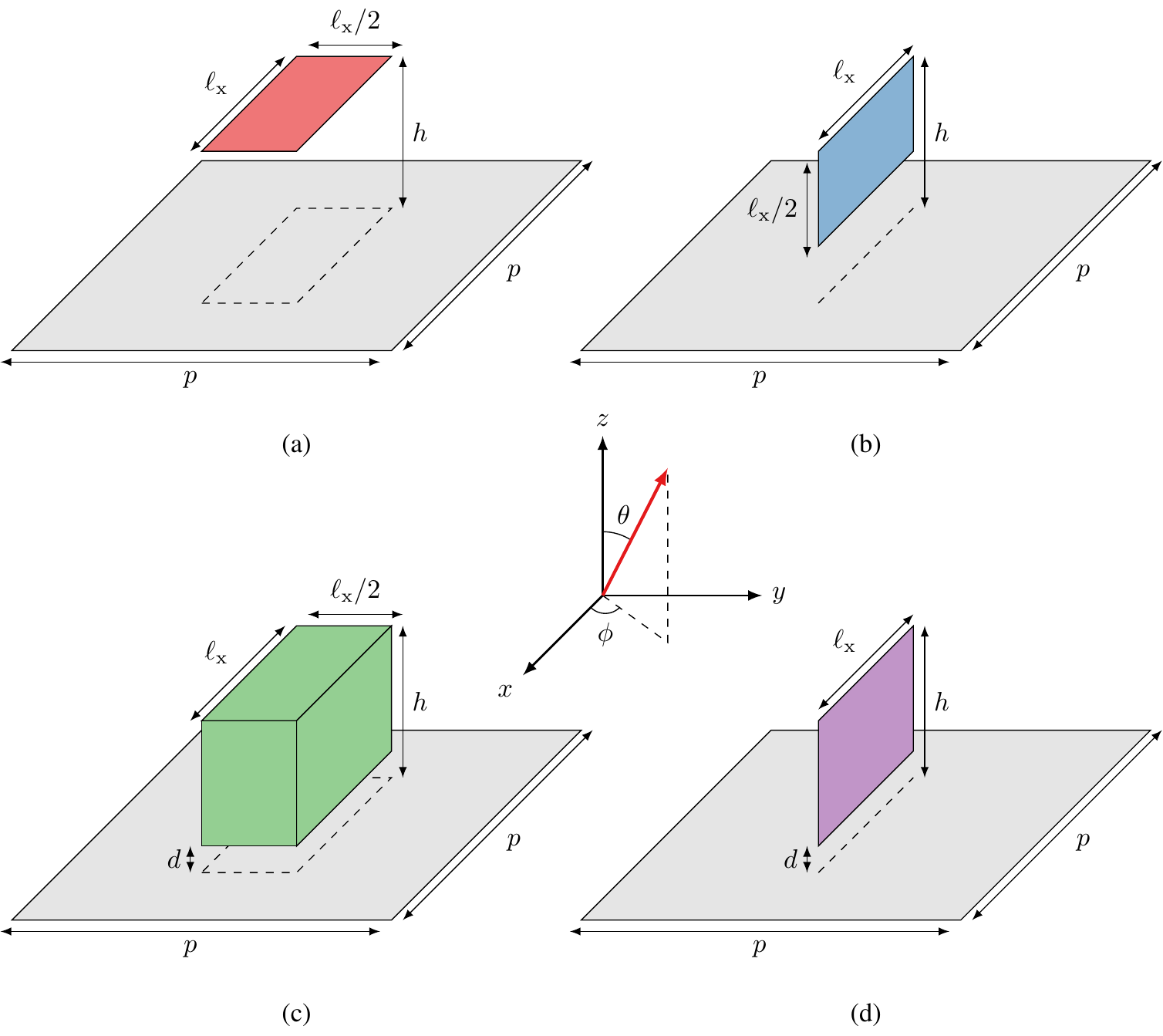}
\caption{Considered array element geometries. (a) Horizontal 2-by-1 plate, (b) Vertical 2-by-1 plate with constant width, (c) Cuboid with constant gap to the ground (d) Vertical plate with constant gap to the ground. Dimensions $p$, $\ell_x$, and $h$ are equal in all four cases. The element area of (a) and (b) is $p^2/9$.}
\label{fig:dimensions}
\end{figure}

\begin{figure}
\centering
\includegraphics{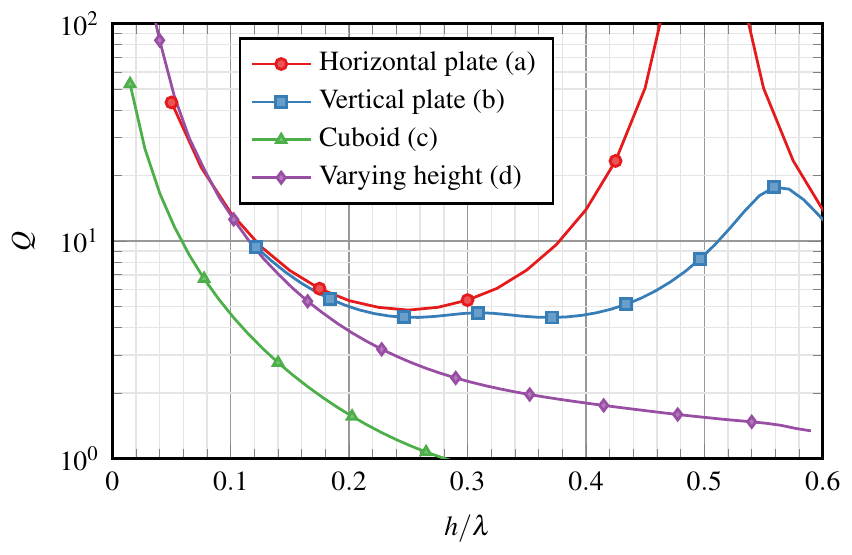}
\caption{Minimum Q at broadside as a function of ground plane distance for the four geometries shown in Fig.~\ref{fig:dimensions}.}%
\label{fig:QvsHeight}
\end{figure}

Fig.~\ref{fig:dimensions} shows three shapes in addition to the horizontal plate: a vertical plate, a cuboid, and a varying height vertical plate. All shapes are over a ground plane. For shape (a) and (c) $\ell_{\rm x}=2\ell_{\rm y}$, $\lx\ly=p^2/9$. Shape (b) has $\ell_{\rm x} = 2\ell_{\rm z}$ and the same area as (a), $d = \frac{1}{24}\lambda$ for (c) and (d). Broadside radiation is considered. The frequency is chosen so that $\lambda=2p$.  The horizontal plate, in Fig.~\ref{fig:QvsHeight}, c.f. Fig.~\ref{fig:Lx_Ly},  has a minimum $Q$-factor at $\frac{1}{4}\lambda$ and  maximum at $\frac{1}{2}\lambda$. 

When the plate is rotated vertically, Fig.~\ref{fig:dimensions}(b), a different result is obtained.
While there still is a minimum at $h =\frac{1}{4}\lambda$ in this case as well, there is another minimum when the lower edge of the plate is at quarter-wavelength height $h\approx 0.37\lambda$. The reason for this, is that the currents are not confined to one height only, as in the horizontal case, and the rapidly increasing Q-factor at $h =\frac{1}{2}\lambda$ does not occur for the vertical plate. There is a more gradual increase when the center of the plate is at a half-wavelength distance from the ground plane.

The Q-factor of the plate with varying height, Fig.~\ref{fig:dimensions}(d), is smaller than the fixed-area vertical plate, Fig.~\ref{fig:dimensions}(b).
From this it is clear that a higher volume usage of the unit cell has a potential to improve the bandwidth.
The thicker cuboid, Fig.~\ref{fig:dimensions}(c), remains below all the other curves in Fig.~\ref{fig:QvsHeight}, providing a larger estimated bandwidth for antenna elements designed to fit within this shape. These two latter results remain fairly insensitive to the distance, $d$, above the ground-plane and only the choice $d=\frac{1}{24}\lambda$ is shown.

\subsection{Scanning in free space}

\begin{figure}[!t]
\centering
\includegraphics{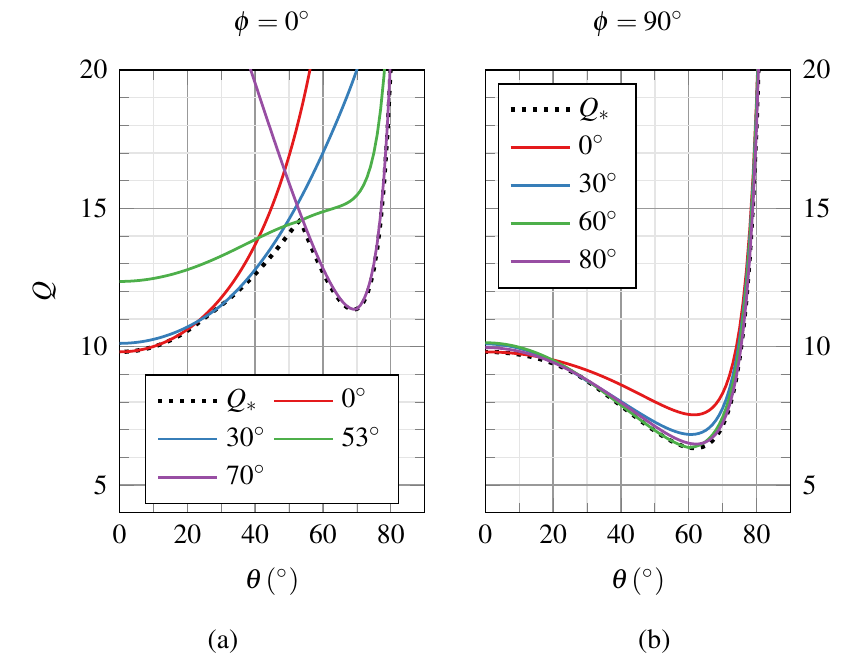}
\caption{Q-factor at different scan angles with no ground plane
and (a) $\phi = 0^\circ$, (b) $\phi = 90^\circ$.
The lowest curve is the optimized Q-factor bound at each scan angle.
The colored curves illustrate the impact on the Q-factor under scanning utilizing a current density associated with the optimal Q-factor for scan angles $\theta$ shown in the legend.
}
\label{fig:scan_angle}
\end{figure}

The results in this section examine the impact of scanning. The scan-direction $(\phi_0,\theta_0)$ appears in $\vec{k}_{\rm t00}$, see \eqref{eq:J_periodic}\, and consequently in $G$ and $g$ \eqref{eq:Greens_function}, \eqref{eq:little_g_exact}, and therefore in stored energies and radiated power.
Here we examine the optimum Q-factor for any antenna element that fits within an array of the horizontal plates each with $\lx = 2\ly$ and $\lx\ly = p^2/9$, and $\lambda=2p$.
Fig.~\ref{fig:scan_angle} shows the minimum Q-factor
as a function of the scan angle $\theta$ in two different planes,
$\phi = 0^\circ$ and $\phi = 90^\circ$, see the coordinate system in Fig.~\ref{fig:dimensions}.
The results are symmetric in each plane crossing the $z$-axis.
The colored curves in Fig.~\ref{fig:scan_angle} correspond to the current of an optimized Q-factor
at the scan angles shown in the legends,
and this current is used to determine the Q-factor for a sweep of scan angles. The dashed curve is the optimal Q-factor at each scan angle.
It is interesting to examine for what range that the local-angle-optimized (colored) graph is similar to the dotted black curve. The range of angles where the Q-factor for the scan-angle specific optimization remain close to optimal is large, in particular for the locally optimized scans at $[0^\circ,30^\circ, 70^\circ]$. 

\begin{figure}[!t]
    \centering
    \includegraphics{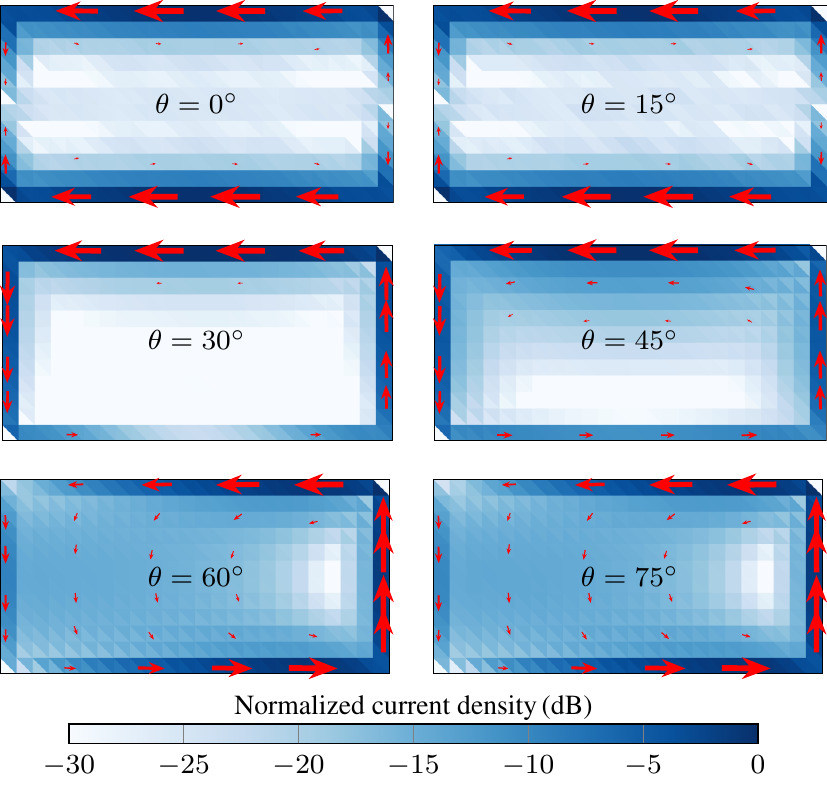}
    \caption{Optimal current distributions for scan angles shown in Fig.~\ref{fig:scan_angle}(a).}%
    \label{fig:scancurrents}
\end{figure}

In the $\phi = 0^\circ$ plane, one can observe that the minimum Q-factor is obtained at broadside, with the Q-factor increasing as the array is steered away.
However, at around $\theta = 53^\circ$ the Q-factor begins to decrease again.
Fig.~\ref{fig:scancurrents} shows the current distributions in the $\phi = 0^\circ$ direction with
$\theta$  shown every 15$^\circ$.
The broadside result shows two dipole-like currents on the long edges of the plate.
For small deviations from broadside, e.g. 15$^\circ$, this current distribution remains similar.
However, as the scan angle is increased, one of the currents on a long edge disappears. The remaining currents extend further into the short edges,
eventually converging on the other long edge to form a loop-like current distribution.
The loop current is imbalanced, remaining stronger on one long edge compared to the other.
After 53$^\circ$, where the Q-factor begins to decrease, the strongest current
can be found on one of the short edges instead.
This current distribution remains stable across the entire $>53^\circ$ scan range.
From the current distributions, it is clear that an optimal current is an interaction between different current modes.

\begin{figure}
    \centering
    \includegraphics{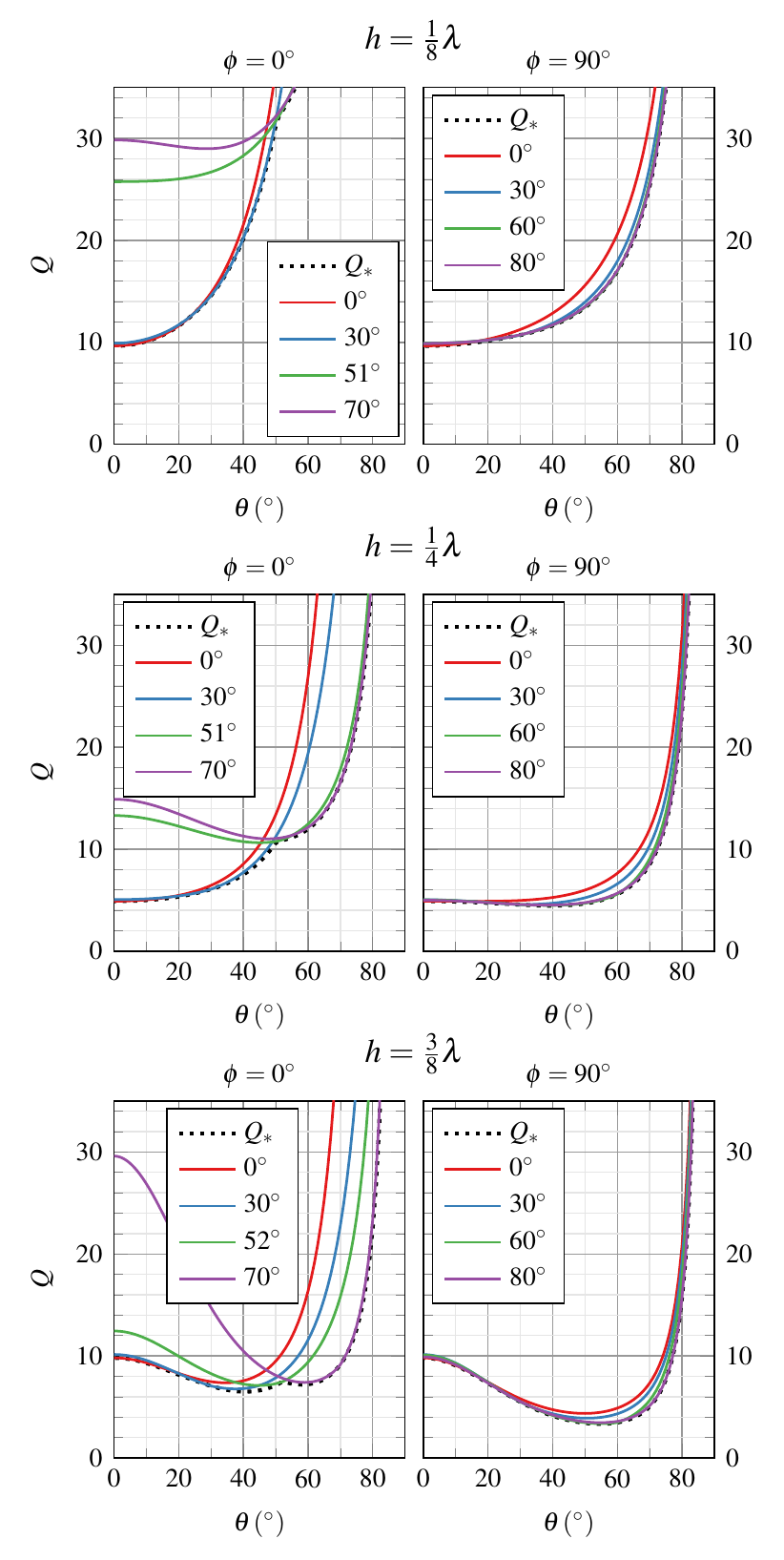}
    \caption{Q-factor as a function of scan angle at three different distances from the ground plane, presented in the same way as in Fig.~\ref{fig:scan_angle}.}\label{fig:scanwgp}
\end{figure}

\subsection{Scanning with a ground plane}

In this section, consider the horizontally oriented rectangular plate with $\ell_x=2\ell_y=p/9$ at height $h$ over a ground-plane at $z=0$. In Fig.~\ref{fig:scanwgp} the same idea as for Fig.~\ref{fig:scan_angle} is repeated, for three different heights, $h=[\frac{1}{8},\frac{1}{4},\frac{3}{8}]\lambda$, above the ground plane. Colored curves are optimized at a given angle, as stated in the legend, the black-dashed curve is the minimum Q-factor at each angle.

From the black-dashed curves for the three different heights, it is clear that for the heights $\frac{1}{8}\lambda$, $\frac{1}{4}\lambda$ and both scan directions, the minimum Q-factors are higher for small $h$ and they tend to increase as the array becomes thinner. In contrast, note that at $h=\frac{3}{8}\lambda$ there is a {\it decrease} in the optimal Q-factor away from the broadside scan direction. Indeed, in both planes for scan-angles above $40^\circ$,
the plate at $\frac{3}{8}\lambda$ outperforms the otherwise superior $\frac{1}{4}\lambda$-case.
This suggests that while the traditional $\frac{1}{4}\lambda$ height
is optimal for narrow-band arrays at broadside, a larger height actually provides the possibility for better scanning performance for small narrow-band elements.

Despite the different values of the Q-factor when compared between different heights and also the case without the ground plane, the current distributions with the ground plane are similar to those shown in Fig.~\ref{fig:scancurrents}. In all cases a similar transition can be seen: lengthening dipole current into a loop, which eventually orients itself on a short edge.
The presence of multiple modes has been utilized to improve scanning in~\cite{Prinsloo2014,hook2010}.
A good way to interpret the above scan results in terms of antenna designs is to think about the optimal current as of the current of a novel antenna shape within the region $\Omega$ that support an optimal current. Such an optimal currents was in~\cite{Jonsson+etal2020} used to design a better antenna for a non-periodic case.

\subsection{Radiation efficiency}

\begin{figure}
\centering
\includegraphics[width=\columnwidth]{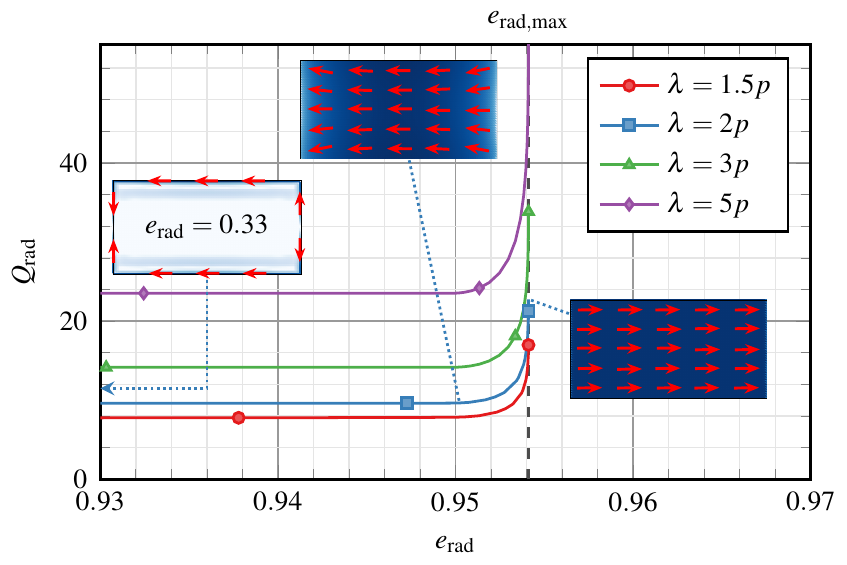}
\caption{Minimum radiation Q as a function of radiation efficiency for an array of rectangular plates in free space and the associated optimal currents. The element configuration is shown in the inset of Fig.~\ref{fig:Lx_Ly} with $\ell_{\rm x}=2\ell_{\rm y}$.}
\label{fig:efficiency}
\end{figure}

In this section, the radiation efficiency constraint is added to the optimization problem.
The minimum $Q_{\rm rad}$ is obtained from the Q-factor optimization without constraints, with current $\Ibf_*$. 
The efficiency obtained with this current using \eqref{eq:P_ohm_constr} is low, as is natural for this choice of $R_s$. Higher efficiency gives an increase in Q-factor. This is implemented by adding a constraint on the dissipation factor \eqref{eq:P_ohm_constr} to \eqref{Qquad}.  However, it does not appear to be a significant increase as long as the efficiency constraint is not pushed too far. For example, in the case of $\lambda = 2p$ the minimum Q-factor of $Q_{\rm rad} \approx 9.57$ is obtained at an efficiency of $\rade\simeq 33\,\%$. Pushing this efficiency to $\rade\simeq 95\,\%$ will have negligible effect on the Q-factor, resulting in $Q_{\rm rad} \approx 9.66$.

The efficiency has an upper bound. For the studied case, this is found at $\rade\simeq 95.4\,\%$. Attempting to reach this bound imposes a high cost on the Q-factor. A similar sharp increase has also been shown in~\cite{Gustafsson+Capek+Schab2019} for single port antennas.
These results are consistent across all studied wavelengths,
the radiation Q-factor grows linearly in $\lambda$ to the first order, e.g.,~$Q_{\rm rad} \propto \lambda$.

The insets of Fig.~\ref{fig:efficiency} illustrate that a demand of higher efficiency results in current densities approaching the uniform distribution to minimize losses.

\subsection{Polarization purity}

\begin{figure}
\centering
\includegraphics{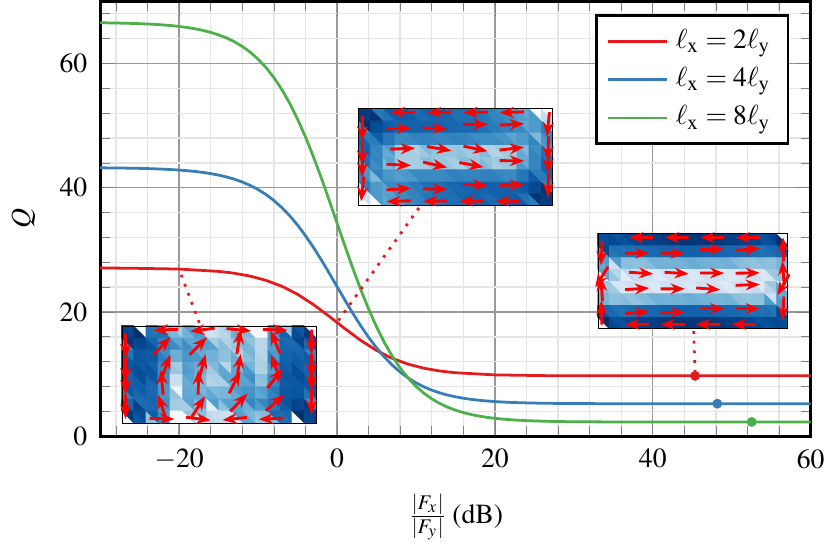}
\caption{Minimum Q as a function of the ratio of x and y-components of the far field for three different aspects ratios of a rectangular plate with fixed area $\lx\ly = p^2/9$ in free space. The dots on the lines denote the minimum Q-factor bounds (with no polarization constraint) and the resulting polarization ratio.}
\label{fig:polarization}
\end{figure}

In this section, a polarization constraint~\eqref{eq:F_constr} is applied to the broadside Q-factor optimization problem. The geometry corresponds to the free-space case, see the inset in Fig.~\ref{fig:Lx_Ly}, with $\lambda = 2p$ and for rectangle aspect ratios: $\ell_{\rm x}/\ell_{\rm y}=[2,4,8]$, and fixed area $\ell_{\rm x}\ell_{\rm y}=p^2/9$. Fig.~\ref{fig:polarization} depicts the minimum Q-factor bound as a function of the ratio of the x- and y-directed polarization components of the Floquet mode \eqref{eq:co+crosspol} for rectangles of different aspect ratios in free space.
Here, let $\vhat{e}_{co}$ and $\vhat{e}_{cx}$ in~\eqref{eq:co+crosspol} be the unit-directions $\vhat{x}$ and $\vhat{y}$ respectively, then $F_x = \vec{F}_{00+}\cdot \vhat{x}$ and $F_y = \vec{F}_{00+}\cdot \vhat{y}$ for the fundamental mode.
The marked circle on each of the lines shows the polarization and Q-factor without any constraints on the polarization. As the studied shapes are longer in the x-direction, the polarization is naturally x-oriented.
Unlike the case with the efficiency constraint, pushing for improved polarization purity does not appear to cause the Q-factor to increase. It remains basically constant for $|F_x|/|F_y|$ from 20 to 60\,dB.
Naturally, forcing the polarization to be aligned with the y-axis causes the Q-factor to increase as the available surface for the y-directed currents is much shorter. This effect gets more pronounced as the aspect ratio of the rectangle is increased.
The insets show the current density as distributed over the element, together with a fixed-time snapshot of the current-directions with unit-length arrows, to improve the visibility.

\section{Conclusions}
\label{sec:conclusions}

This paper presents two optimization methods to obtain Q-factor-based estimates of bandwidth for phased array antennas.
The examples demonstrate the usefulness and the accuracy of such estimates for small narrow-band antenna elements. The paper investigates certain array element form-factors, beam-scanning, placement of the element relative to the ground plane, a trade-off between the radiation efficiency and the Q-factor, and a trade-off between the polarization purity and the Q-factor.
 
The proposed optimization methods efficiently solve the originally non-convex problems of Q-factor minimization
and determine an optimal current density corresponding to the minimum value.
Such optimal currents can provide insight into the physics behind optimal narrow-band radiation, as for example in  Fig.~\ref{fig:Lx_Ly} where
the distributions indicate that certain double-dipole type currents tend to minimize the Q-factor.
Similarly, a change in current distribution illustrates the existence of different types of current modes; at broadside, and  at high-scan angles for antennas within the investigated narrow-band rectangular plate-shapes.

Another interesting result is the bandwidth comparison between realized antenna elements and the Q-factor bounds. Here it is shown that, for a family of elements at different heights above the ground, the Q-factor well predicts the available bandwidth of narrow-band elements, e.g.,  $Q\gtrsim 5$.
For more broadband elements, a single Q-factor resonance tends to underestimate the available bandwidth. These Q-factor bounds and their associated bandwidth estimates are interesting since they are both easy to obtain, and more predictive than the array-figure of merit bound. The latter is useful for wide-band multi-resonant antenna elements, where it provides the absolute upper bound for lossless arrays. The Q-factor bound gives information on the required size, scanning impact, and placement within the unit-cell.

In this paper, we have also investigated the behavior of optimal Q-factor for any antenna in a given region for a given scan direction, both with and without a ground plane. It was especially interesting to note that for an optimal antenna in these small enclosing regions there exists a different height (above $\frac{1}{4}\lambda$) that has a better Q-factor performance at large scan ranges.
The two constrained Q-factor optimization cases, efficiency and polarization purity show the versatility of the Q-factor approach in estimating the bandwidth.

The above examples illustrate that the Q-factor optimization of periodic structures is of significant interest. It is more predictive in estimating the available bandwidth  for narrow-band structures than e.g. a sum-rule. Indeed, physical insight can be obtained in certain cases, such as the classical quarter-wavelength-position for optimal antennas above a ground plane. It is also easy to add constraints to the optimization problem, such as  polarization purity. In addition, Q-factor optimization provides trade-off relations where e.g., the height or aspect ratio can be tuned at the expense of the Q-factor. The higher bandwidth robustness of a vertically oriented rectangle in comparison with a horizontally oriented rectangle is clearly visible in the Q-factor of this case. 

\appendix

\section{Appendix}
\label{sec:appendix}

The free-space two-dimensional periodic Green's function in the spectral form is~\cite{Oroskar+etal2006} 
\begin{equation}
\begin{split}
\G&(\vec{r}_1,\vec{r}_2) 
=\frac{1}{2\iu ab} 
\underset{(m,n)\in \Z^2}{\sum}
\frac{1}{\Kzmn}\eu^{-\iu\Ktmn\cdot(\vec{\rho}_1-\vec{\rho}_2)}\eu^{-\iu \Kzmn|z_1-z_2|},
\label{eq:Greens_function}
\end{split}
\end{equation}
with $\Ktmn=\Ktnull+2\pi\frac{n}{a}\hat{\vec{x}}+2\pi\frac{m}{b}\hat{\vec{y}}$, $\Kzmn=\sqrt[]{k^2-\Ktmn\cdot \Ktmn}$, and $z_i=\vr_i\cdot\vhat{z}$, $\vec{\rho}_i= \vr_i - \hat{\vec{z}}z_i$, $i=\{1,2\}$.

The function $g$ in stored energies kernels~\cite{LudvigOsipov+Jonsson2019} is  
\begin{multline}
g(\vec{r}_1,\vec{r}_2) = 
\frac{1}{4 ab} 
\underset{(m,n)\in \Z^2\setminus \mathcal{P}}{\sum}
\frac{1}{|\Kzmn|^2} 
\eu^{\iu \Ktmn\cdot(\vec{\rho}_1-\vec{\rho}_2)} \\  \eu^{-|\Kzmn||z_2-z_1|}  \left( \frac{1}{|\Kzmn|} + |z_1-z_2|  \right),
\label{eq:little_g_exact}
\end{multline}
where $\mathcal{P}=\{(m,n):k^2 - \Ktmn\cdot\Ktmn\geq 0\}$.

\bibliography{stored}
\bibliographystyle{IEEEtran}

\begin{IEEEbiography}[{\includegraphics[width=1in,height=1.25in,clip,keepaspectratio]{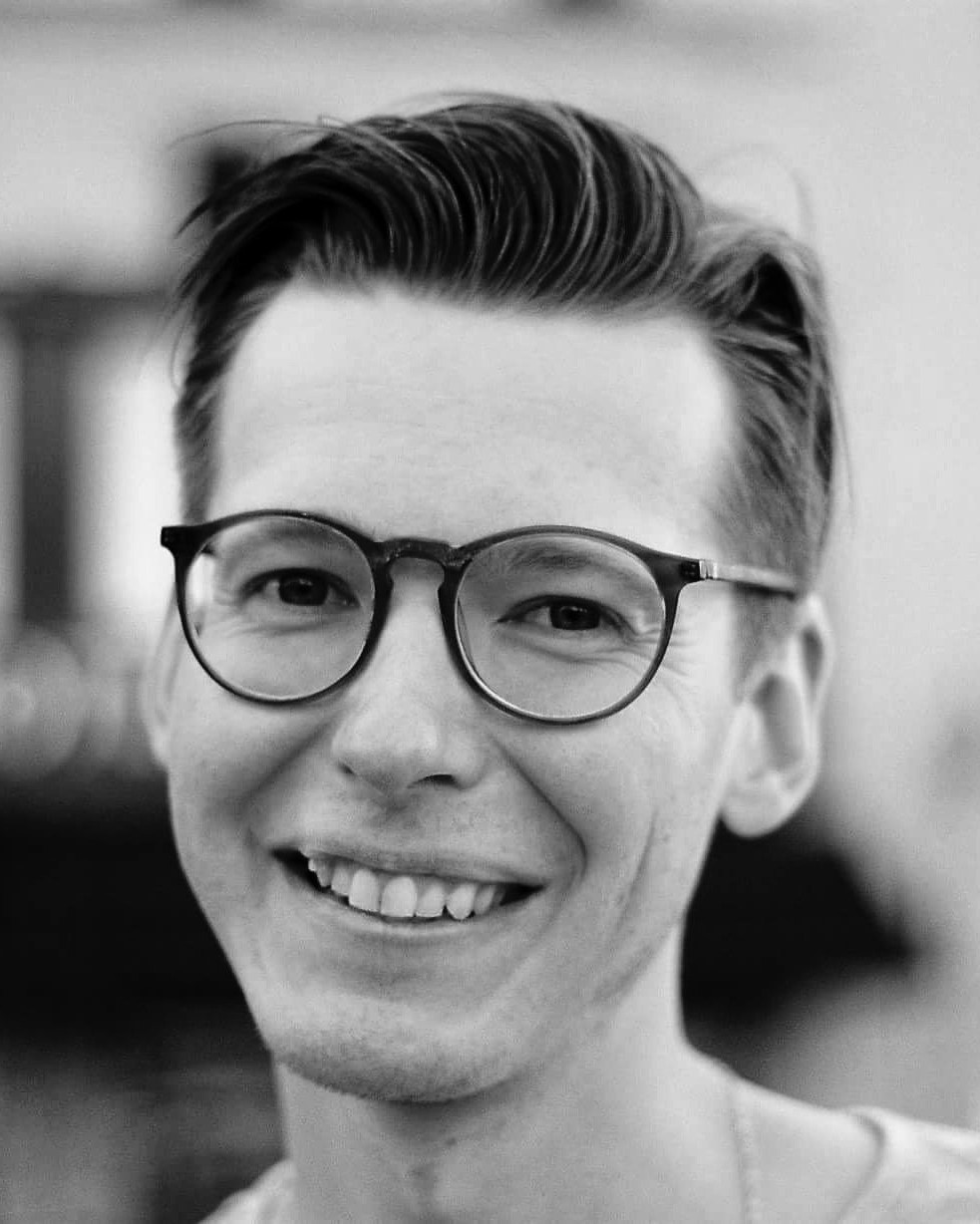}}]{Andrei Ludvig-Osipov} received his B.Sc. and M.Sc. degree in radiofrequency engineering and telecommunications from St.~Petersburg State Electrotechnical University, Russia, in 2012 and 2014 respectively. 
	He received the Ph.D. degree in electrical engineering from KTH Royal Institute of Technology, Stockholm, Sweden in 2020.  
	Since 2020 he is a Postdoctoral Researcher at Chalmers University of Technology, Gothenburg, Sweden. 
He was a research engineer at R\&D Institute of Radio Science and Telecommunications, St.~Petersburg, Russia in 2013-2015.
His research interests include electromagnetic theory, its applications in engineering (antenna design tools and methods) and in physics (plasma and particle physics), signal processing, and numerical simulation methods.
\end{IEEEbiography}

\begin{IEEEbiography}[{\includegraphics[width=1in,height=1.25in,clip,keepaspectratio]{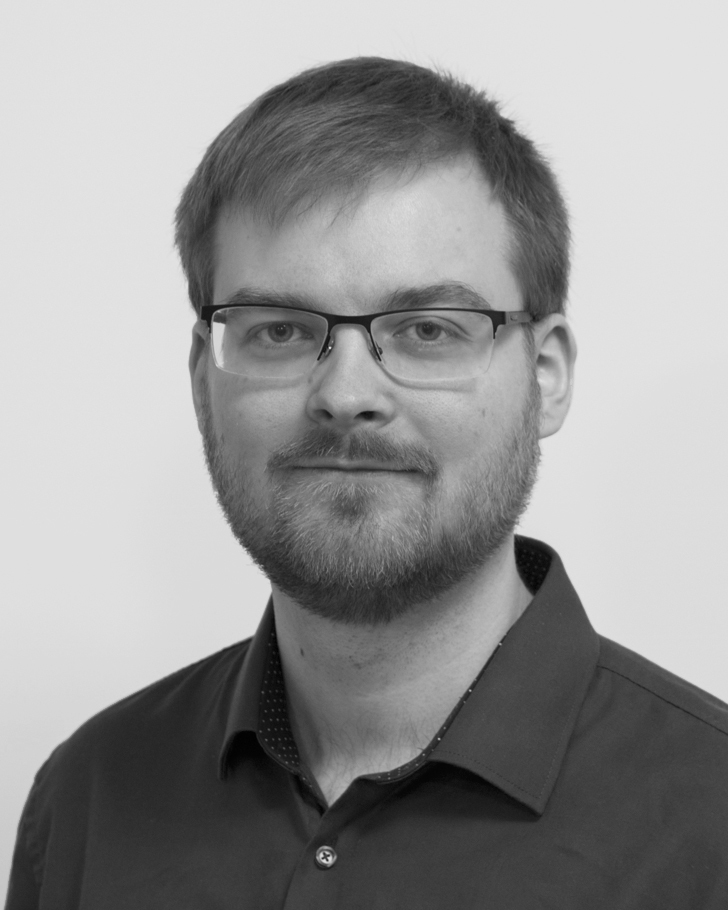}}]{Jari-Matti Hannula} (S'16--M'18) was born in Luvia, Finland, in 1990. He received the B.Sc.\ (Tech.), M.Sc.\ (Tech.), and D.Sc.\ (Tech.) degrees in electrical engineering from Aalto University, Espoo, Finland, in 2014, 2015, and 2018, respectively.

From 2013 to 2019 he was at Aalto University. Since 2019 he has been at KTH Royal Institute of Technology, Stockholm, Sweden, as a Postdoctoral Researcher. His research interests include antenna optimization, multiport antennas, and antenna-transceiver codesign.
\end{IEEEbiography}

\begin{IEEEbiography}[{\includegraphics[width=1in,height=1.25in,clip,keepaspectratio]{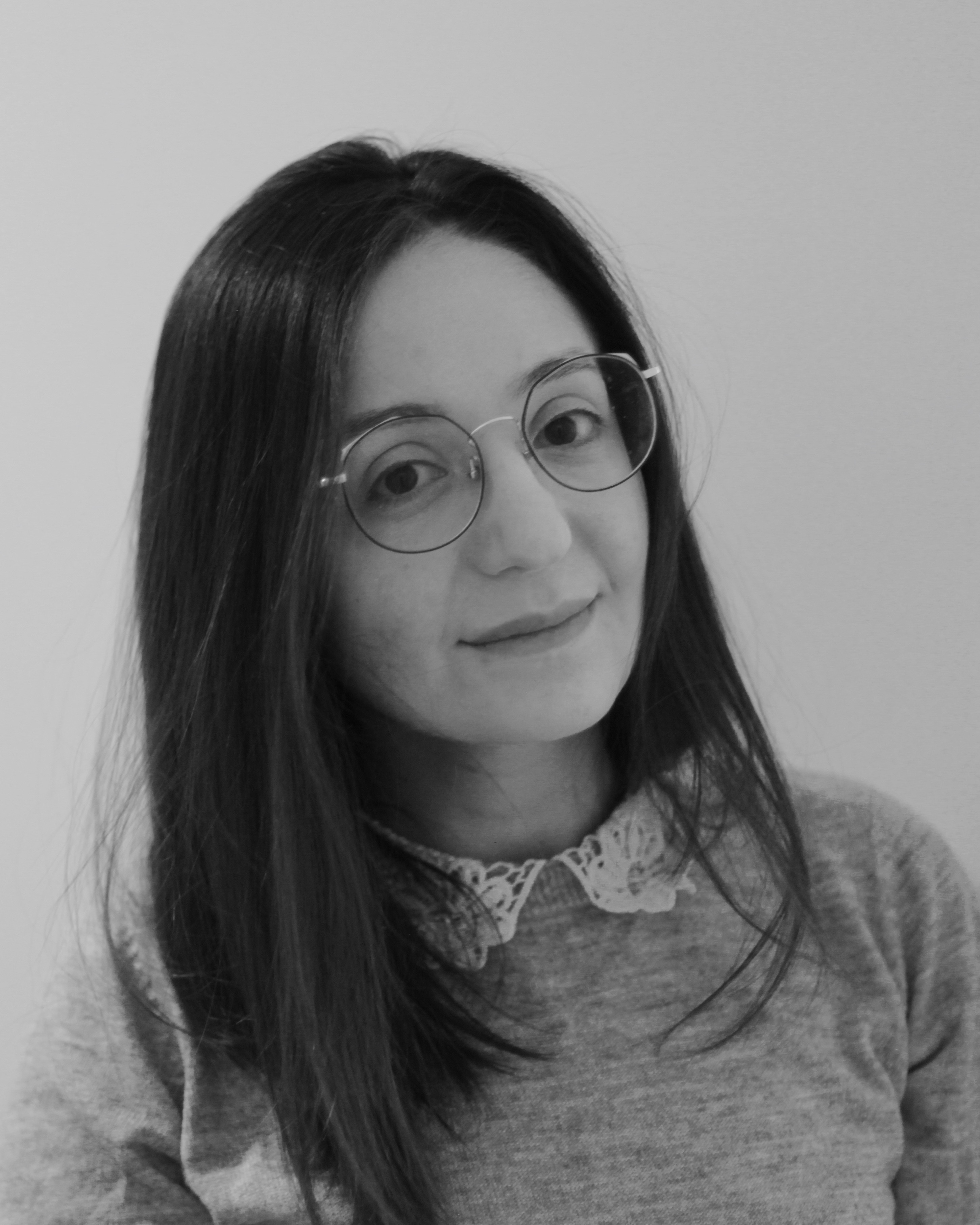}}]{Patricia Naccachian} received her B.E. degree in Electrical and Computer Engineering in 2020 from the American University of Beirut, Lebanon. 
	
Since February 2020, she has been a research assistant intern at KTH Royal Institute Of Technology, Stockholm, Sweden. Her research interests include radio frequency energy harvesting, antenna theory and design, natural language processing, and artificial intelligence. 
\end{IEEEbiography}

\begin{IEEEbiography}[{\includegraphics[width=1in,height=1.25in,clip,keepaspectratio]{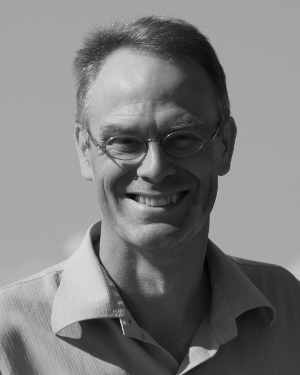}}]{B. L. G. Jonsson} received his Ph.D. degree in electromagnetic theory in 2001 from KTH Royal Institute of Technology, Stockholm, Sweden. He was a postdoctoral fellow at University of Toronto, Canada and a Wissenschaftlicher Mitarbeiter (postdoc) at ETH Zurich, Switzerland. Since 2006 he is with the Electromagnetic Engineering Lab at KTH. He is professor in Electromagnetic fields at KTH since 2015. His research interests include electromagnetic theory in a wide sense, including scattering, antenna theory and nonlinear dynamics. 
\end{IEEEbiography}
\vfill
\end{document}